\documentclass[aps,pra,twocolumn,showpacs,superscriptaddress]{revtex4-1}
\usepackage{hyperref}
\usepackage{amsmath}
\usepackage{bm}
\usepackage{amsfonts}
\usepackage{amsthm}
\usepackage{amssymb}
\usepackage{ifthen}
\usepackage{color}
\usepackage{graphicx}
\newboolean{qcircuit}

\setboolean{qcircuit}{true}     

\newcommand{\ket}[1]{\left |  #1 \right \rangle}

\newcommand{\ba}{\begin{eqnarray}}
\newcommand{\ea}{\end{eqnarray}}

\definecolor{nblue}{rgb}{0.2,0.2,0.7}

\newcommand{\Na}{\mathcal{N}_\text{a}}
\newcommand{\Nb}{\mathcal{N}_\text{b}}
\newcommand{\Nc}{\mathcal{N}_\text{c}}

\begin{document}

\title{Randomness in post-selected events}

\author{Le Phuc Thinh}
\affiliation{Centre for Quantum Technologies, National University of Singapore, 3 Science Drive 2, Singapore 117543}
\author{Gonzalo de la Torre}
\affiliation{ICFO-Institute of Photonic Sciences, Mediterranean Technology Park, 08860 Castelldefels (Barcelona), Spain}
\author{Jean-Daniel Bancal}
\affiliation{Centre for Quantum Technologies, National University of Singapore, 3 Science Drive 2, Singapore 117543}
\author{Stefano Pironio}
\affiliation{Laboratoire d'Information Quantique, Universit\'e Libre de Bruxelles (ULB), Brussels, Belgium}
\author{Valerio Scarani}
\affiliation{Centre for Quantum Technologies, National University of Singapore, 3 Science Drive 2, Singapore 117543}
\affiliation{Department of Physics, National University of Singapore, 2 Science Drive 3, Singapore 117542}

\begin{abstract}
Bell inequality violations can be used to certify private randomness for use in cryptographic applications. In photonic Bell experiments, a large amount of the data that is generated comes from no-detection events and presumably contains little randomness. This raises the question as to whether randomness can be extracted only from the smaller post-selected subset corresponding to proper detection events, instead of from the entire set of data. This could in principle be feasible without opening an analogue of the detection loophole as long as the min-entropy of the post-selected data is evaluated by taking all the information into account, including no-detection events. The possibility of extracting randomness from a short string has a practical advantage, because it reduces the computational time of the extraction.

Here, we investigate the above idea in a simple scenario, where the devices and the adversary behave according to i.i.d. strategies. We show that indeed almost all the randomness is present in the pair of outcomes for which at least one detection happened. We further show that in some cases applying a pre-processing on the data can capture features that an analysis based on global frequencies only misses, thus resulting in the certification of more randomness. We then briefly consider non-i.i.d strategies and provide an explicit example of such a strategy that is more powerful than any i.i.d. one even in the asymptotic limit of infinitely many measurement rounds, something that was not reported before in the context of Bell inequalities.
\end{abstract}

\maketitle

\section{Introduction}

Sources of randomness have numerous applications: in algorithms, samplings, numerical simulations, gambling, and of course cryptography~\cite{MU49,MR95,vadhan11}. The last application demands sources that can be certified as being uncorrelated to any outside process or variable, i.e. private randomness. Typically, the output of a physical process (thermal noise, shot noise, ...) is considered random in this sense only if certain assumptions are made on its underlying behavior. The violation of Bell inequalities, however, certifies private randomness in a device-independent way~\cite{colbeckThesis,PAM10}. From the amount of violation, one obtains a lower bound on the min-entropy $H$ of the output string generated by the process~\cite{PAM10,PM11,VV12}. This information is then sufficient to extract randomness: indeed, one can design seeded extractors, whose output is a string of (roughly) $H$ bits guaranteed to be uniformly random, even according to an external adversary. 

A Bell experiment, however, produces much more information than the mere violation of a single inequality. For instance, one can estimate the single-run frequencies $p(a,b|x,y)$ of the outcomes $(a,b)$ conditioned on the settings $(x,y)$. When this knowledge is taken into account, higher values for the lower bounds on $H$ can in principle be obtained~\cite{bancal13more,silleras14more}. More generally, there may be other ways to process the data that can lead to improved bounds on the randomness, as the following example illustrates.

Consider a Bell experiment running for two days, each day consisting of $N\gg 1$ runs. Suppose that, on the first day, the setup produces outcomes that violate the CHSH inequality maximally; on the second day, for some technical glitch, the detectors don't fire, so the list of outcomes consists only of double no-detection events. Suppose that the users estimate the amount of randomness generated using solely the observed CHSH violation $I$, using the simple bound $H\geq 1-\log_2\left(1+\sqrt{2-I^2/4}\right)$ \cite{PAM10}. Suppose further that they planned to extract randomness every two day. Over the two day period, they observe an average CHSH violation of $(2\sqrt{2}+2)/2\simeq 2.41$ (we take the convention that no-detection events are mapped to $+1$ outcomes), from which they deduce a randomness rate of $\sim 0.2$ bit/run for Alice's outcomes, that is $\sim 0.4N$ bits in total for the two-day period. However, the users might have chosen to extract randomness at the end of each day instead. The same techniques certify now $1$ bit/run for Alice on the first day and 0 on the second, for a total of $N$ bits over the two days \footnote{Notice that the difference grows linearly with $N$, thus it cannot be accounted for by finite-size corrections related to processing two $N$-symbol sets instead of a single $2N$-symbol one.} What happened is clear: the data contain the information that two processes are involved; this information was missed by the overall analysis, but was revealed by the choice of sorting the data in two blocks.

The example is extreme, but a simple variation is very relevant: the case in which no-detection events are evenly spread during the whole duration of the experiment is a good approximation to the data produced in photonics Bell tests, in which no-detection events constitute a large fraction of the runs (see e.g. Table I in \cite{christensen13}). No-detection events come from two processes: the finite efficiency of the detectors, and the fact that parametric down-conversion often produces the vacuum state. The physics of both suggests that these events contain little or no randomness: it is thus tempting to sort the outcomes of the Bell test in two groups, the detections and the no-detections. As in the previous example, this may lead to certify more randomness. Even if it does not, one may get a \textit{practical} advantage by extracting randomness only from the detection events. Indeed, randomness extractors require an independent random seed: the longer the initial string, the longer the needed seed and the computational time to output the result; in fact, it is an active research direction to construct randomness extractor with short seed length~\cite{vadhan11}. Thus, it is beneficial to be able to extract randomness from a short string.

Here, we investigate the amount of randomness that can be certified in Bell tests within the subset of detection events.  For this first study, our aim is simply to determine whether this is actually a viable strategy. We thus perform our analysis in the simplified scenario in which the devices and the adversary behave in an i.i.d. way and in the limit of infinitely many measurement rounds. If randomness cannot be certified in this simple scenario, then it can also certainly not be certified in the non-i.i.d. finite statistics case. 

The post-selection of detection events notoriously opens the detection loophole~\cite{eberhard93,mermin86}. It is important to clarify that our approach does not fall into that trap. We shall compute a lower bound on the randomness that can be extracted from a subset of events, but the bound is obtained \textit{by taking into account the whole set of events}. In particular, if the behavior of the devices is compatible with local realism due to the detection loophole, our method will say that no randomness can be certified in the post-selected set of detection events.

Let us remark that a similar analysis in the context of violating local realism, namely the p-value of post-selected events which does not contain no-detections, has been done recently~\cite{B15}.

After introducing the technique that we will use to bound randomness in Section~\ref{sec:method}, we apply it to several physically-motivated examples in section~\ref{sec:examples}. In Section~\ref{sec:perfectDet} we analyse more precisely the effect of post-selection in a simplified case. A glimpse beyond the i.i.d. restriction is given in Section~\ref{sec:noniid} before the conclusion.

\section{Average randomness in post-selected events}\label{sec:method}

Consider a Bell experiment consisting of two separate devices in which each party inputs $x\in\mathcal{X}$ and  $y\in\mathcal{Y}$ and obtains outputs $a\in\mathcal{A}$ and $b\in\mathcal{B}$, respectively. The behavior of such devices over $n$ successive runs can be characterized by the -- generally unknown -- joint probabilities $p(\mathbf{ab}|\mathbf{xy})$ to obtain the output string $\mathbf{ab}=(a_1b_1,\ldots,a_nb_n)$ given the input string $\mathbf{xy}=(x_1y_1,\ldots,x_ny_n)$. The information that an adversary has over the output string can be characterized by a tripartite quantum distribution $p(\mathbf{abe}|\mathbf{xyz})$ where $\mathbf{e}$ denotes the output the adversary obtains when he makes a measurement $\mathbf{z}$ on a system possibly entangled with Alice and Bob's devices. In general $\mathbf{e}$ can be a string of arbitrary size representing the total information that the adversary can get about Alice and Bob's outcomes and $\mathbf{z}$ can be an arbitrary measurement that depends on the information available to the adversary in the protocol before his measurement.

Here we shall make the following simplifying assumptions. First, we will assume that the device behave in an i.i.d. way and similarly that the adversary extracts his information in an i.i.d. way by performing at each run individual measurements $z_i$. We can thus write $p(\mathbf{abe}|\mathbf{xyz})=\prod_{i=1}^n p(a_ib_ie_i|x_iy_iz_i)$. Second, we are going to assume that Alice and Bob's marginal $p(ab|xy)$ at each run are known and given. In this way, we do not need to take care of estimation. With these assumptions, finding the adversary's optimal attack thus amounts at optimizing some quantity over all tripartite quantum distributions $p(abe|xyz)=\langle  \Psi|M_{a|x}\otimes M_{b|y}\otimes M_{e|z}|\Psi\rangle$ compatible with a given bipartite marginal $p(ab|xy)=\sum_e p(abe|xyz)=\langle  \Psi|M_{a|x}\otimes M_{b|y}\otimes I|\Psi\rangle$.

Let us now introduce the additional ingredient of post-selection. For this, we consider a bipartition of the joint output alphabet $\mathcal{O}=\mathcal{A}\times\mathcal{B}$ into two sets $\mathcal{V}$ (valid symbols) and $\mathcal{N}$. If the outputs at a given round $(a,b)\in\mathcal{V}$, we say that the round is valid, and otherwise, if $(a,b)\in\mathcal{N}$, that it is invalid. We refer to the events obtained in valid runs only as the post-selected events. Our goal is to estimate how much randomness can be extracted from these post-selected events. 

A priori, an adversary trying to guess the post-selected events might not have access to the information about which run turned out to be valid or invalid, since he should not have access to the outputs observed by the parties. For simplicity, however, we'll assume here that the adversary has access to this information. This allows him to know exactly which run he should try to guess and is thus advantageous for him. The amount of randomness that can be certified in this case thus constitutes a lower bound on the amount that can be certified when the adversary is not given this information. This assumption might however be problematic in a non-i.i.d. situation (see Section~\ref{sec:noniid}).

We are going to assume in the following that Alice and Bob use a certain pair of inputs ($\bar x$,$\bar y$) for randomness generation \footnote{In the case they would rather use several pairs of inputs for randomness generation, the analysis below could be extended by using the tools presented in~\cite{bancal13more}.}. Since there is a promise on the marginal $p(ab|xy)$ and since we do not need to consider how to estimate this quantity, we are going to assume for simplicity that Alice and Bob always measure their systems using the inputs ($\bar x$,$\bar y$). Suppose that by measuring $n$ systems, they obtain $m$ results in $\mathcal{V}$ and $n-m$ results in $\mathcal{N}$. The number $m$ of valid results is a random variable with probability distribution $p(m)={n \choose m} p_{\bar x \bar y}^m(1-p_{\bar x \bar y})^{n-m}$, where $p_{\bar x \bar y}=\sum_{ab\in\mathcal{V}}p(ab|\bar x\bar y)$ is the single-run probability to obtain a pair of valid results when using inputs $(\bar x,\bar y)$.

By the i.i.d. assumption, the min-entropy of the $m$-elements post-selected string is $m\, H_{\bar x \bar y}$, where $H_{\bar x \bar y}$ is the single-run min-entropy and is defined below. Applying a randomness extractor to this string, then yields $m\, H_{\bar x \bar y}$ bits of randomness (such extractors exist up to $\epsilon$ correction, see \cite{shaltiel2002recent,DPVR12}). The average length of the final random string is then $\sum_{m=0}^n p(m) m\, H_{\bar x \bar y} = n\,p_{\bar x \bar y}\,H_{\bar x \bar y}$. We can also interepret this last quantity as an ``average" min-entropy \footnote{Note that usually, the average min-entropy of a variable $A$ (e.g. the output string) given some information $M$ (e.g. the length of the post-selected string) is defined as $-\log_2 \sum_m P(m) G(A|M=m)$ where $G(A|M=m)$ is the guessing probability of $A$ given $M=m$. Here our definition of ``average'' min-entropy is $\sum_m P(m)(-\log_2 G(A|M=m))$, that is, we inverted the sum over $m$ and the logarithm. The reason is that in our scenario, the user, and not only the adversary, actually knows the value of $m$ and thus a bound on $G(A|M=m)$ which allows him to apply a different extractor depending on $m$.}. The rate of randomness extraction per use of the device can then be defined as $p_{\bar x \bar y}\, H_{\bar x \bar y}$.

To complete the analysis, it remains to determine $H_{\bar x \bar y}$. By definition, the min-entropy is related to the guessing probability $G_{\bar x \bar y}$ as $H_{\bar x \bar y}=-\log_2 G_{\bar x \bar y}$, where the guessing probability is the maximal probability that the adversary correctly guesses Alice and Bob's outputs by performing an optimal measurement on his quantum side information \cite{KRS09}. Here since we condition on valid runs, this quantum side information can be represented by the cq-state $\rho_{ABE}=\frac{1}{p_{\bar x \bar y}}\sum_{ab\in\mathcal{V}}|ab\rangle\langle ab|\otimes \rho_{E}^{ab}$, where $\rho_{E}^{ab}=\text{tr}\left(M_{a|\bar x}\otimes M_{b|\bar y}\otimes I\, |\Psi\rangle\langle\Psi|\right)$. The probability that the adversary then makes a correct guess $e=(a,b)$ of Alice and Bob's outputs $a,b$ by performing a measurement $z$ on his system is, averaged over Alice and Bob's possible outputs, $\frac{1}{p_{\bar x \bar y}}\sum_{ab\in\mathcal{V}} \text{tr}\left(M_{ab|z} \rho_E^{ab}\right)=\frac{1}{p_{\bar x \bar y}}\sum_{ab\in\mathcal{V}} \langle \Psi|M_{a|\bar x}\otimes M_{b|\bar y}\otimes M_{ab|z}|\Psi\rangle$. To determine the maximal value of this guessing probability, we should maximize it over all quantum realizations $R=(|\Psi\rangle,\{M_{a|x}\},\{M_{b|y}\},\{M_{e|z}\})$ compatible with the given marginals $p(ab|xy)$ characterizing Alice and Bob's devices. We thus have
\begin{eqnarray}
G_{\bar x \bar y} = \frac{1}{p_{\bar x \bar y}} &\underset{R}{\max} & \sum_{ab\in\mathcal{V}} \langle \Psi|M_{a|\bar x}\otimes M_{b|\bar y}\otimes M_{ab|z}|\Psi\rangle\label{goptim1}\\
& \text{s.t.} & \langle \Psi|M_{a|x}\otimes M_{b|y}\otimes I|\Psi\rangle=P(ab|xy)\,.\nonumber
\end{eqnarray}

Following \cite{silleras14more} and introducing the bipartite subnormalized quantum correlations $\tilde p_{a'b'}(ab|xy)=\langle\Psi|M_{a|x}\otimes M_{b|y}\otimes M_{a'b'|\bar z}|\Psi\rangle$ where $\bar z$ denotes the adversary's optimal measurement which maximizes (\ref{goptim1}), the above optimization program can be rewritten as
\begin{eqnarray}
G_{\bar x \bar y} = \frac{1}{p_{\bar x \bar y}} &\ \underset{\tilde p_{a'b'}}{\max}\ & \sum_{ab\in\mathcal{V}} \tilde p_{ab}(ab|\bar x\bar y)\nonumber\\
& s.t. & \sum_{a'b'\in\mathcal{V}}  \tilde p_{a'b'}(ab|xy)=p(ab|xy)\label{goptim}\\
& & \tilde p_{a'b'}(ab|xy) \in \tilde Q\nonumber
\end{eqnarray}
where $\tilde Q$ denotes the set of unormalized bipartite quantum correlations. The meaning of this program is intuitive: Eve prepares one of $|\mathcal{V}|$ systems for Alice and Bob, one for each outcome pair $(a'b')$. Each system is characterized by joint probabilities $p_{a'b'}(ab|xy)=\tilde p_{a'b'}(ab|xy)/q_{a'b'}$ and is prepared with probability $q_{a'b'}=\sum_{ab}\tilde p_{a'b'}(ab|xy)$. When Eve prepares system $ab$, she guesses that Alice's and Bob's outputs are $ab$, hence the probability that she guesses correctly on average is given by the objective function in (\ref{goptim}). Eve's preparations should of course on average reproduce the given correlations $p(ab|xy)$, hence the first constraint of (\ref{goptim}). The second constraint simply expresses that Eve's preparations should be compatible with quantum theory.

Notice that the constraints in the second line of (\ref{goptim1}) and in the second one of (\ref{goptim}) involve all outputs $a,b$ and not only those belonging to the post-selected set $\mathcal{V}$. This reflects the fact that our analysis is not subject to the detection loophole.

To summarize, for a given set of bipartite correlations $p(ab|xy)$ characterising the behavior of the devices, the figure of merit that we are going to consider in this paper, which we call the randomness rate, is $p_{\bar x\bar y} H_{\bar x\bar y}=p_{\bar x\bar y}\times(-\log_2 G_{\bar x\bar y})$ where $G_{\bar x\bar y}$ is the output of the optimization problem (\ref{goptim}).

In general, it is not possible to carry out explicitly this optimization as there is no closed form for the set of quantum correlations $\tilde Q$. However, we can upper-bound the optimal value of (\ref{goptim}), and thus lower-bound the randomness rate, through semidefinite programming by relaxing the last condition $\tilde P_{a'b'}(ab|xy) \in \tilde Q$ and asking that $\tilde P_{a'b'}(ab|\bar x\bar y)$ belongs to some level of the NPA hierarchy~\cite{qtest0,qtest,qtest2} instead of the exact quantum set.
All optimizations reported here were performed at local level 1 of the SDP hierarchy~\cite{moroder2013}.

\section{Approximating photonic experiments\label{sec:examples}}
The natural benchmark to test our tools are the correlations expected in a Bell experiment using spontaneous parametric down-conversion (SPDC). In the single-mode case, such a pulsed SPDC source produces a state of the form
\begin{equation}
\ket{\psi}=c(g,\bar g)e^{\tanh(g) a_H^\dagger b_V^\dagger - \tanh(\bar g)a_V^\dagger b_H^\dagger}\ket{0},
\end{equation}
where $a_{H/V}$ $(b_{H/V})$ are polarization modes for Alice (Bob), $\ket{0}$ is the vacuum state, and $c(g,\bar g)=\sqrt{1-\tanh^2 g}\sqrt{1-\tanh^2 \bar g}$ for $g,\bar g$ being the two squeezing parameters. The parties Alice and Bob can measure this state by placing two detectors after the usual set of wave plates and a polarization beam splitter. If the detectors do not resolve the number of incident photons, four cases can then be observed: no detection, a click in the first detector, a click in the second detector, or two clicks. In the following, we label a click in the first detector as 0, a click in the second detector as 1, and the case where either no detection or double detections are observed as $\varnothing$, so that each party effectively produces one of three possible outcomes. The statistics observed in this situation as a function of the polarization measurements and the detection efficiency (or equivalently the losses between the source and the detectors) are described in~\cite{VSBLCMTZGS15}. 

Using the program~\eqref{goptim}, we are going to compute lower bounds on the extractable randomness that can be found in presence of these statistics in the following cases:

\begin{itemize}
\item (a) All outcomes are considered (no post-selection), i.e. $\mathcal{N} = \Na = \{\}$ (the empty set).
\item (b) The post-selected string of outcomes does not contain double occurrences of $\varnothing$, i.e. $\mathcal{N}=\Nb=\{\varnothing\varnothing\}$.
\item (c) The post-selected string of outcomes does not contain any occurrence of a no-detection event $\varnothing$, i.e. $\mathcal{N}=\Nc=\{0\varnothing$, $1\varnothing$, $\varnothing\varnothing$, $\varnothing0$, $\varnothing1\}$.
\end{itemize}


For the sake of comparison, we will sometimes also consider the case in which the measurements are performed only when at least one photon pair is produced by the source, i.e.
\begin{itemize}
\item (h) The source is heralded.
\end{itemize}
An example of heralded experiment is the recent one of Hensen et al.~\cite{HBDRKBRVSAAPMMTEWTH15}. Note that in this particular case the state is encoded in a non-photonic system and always yields a detection whenever measured.

\subsection{Perfect detectors, variable squeezing}

We first consider the case of an experiment with no loss, and with unit efficiency detectors. In this case it seems natural to try to generate a maximally entangled state. We thus set $g=\bar g$ and vary the squeezing $g$. Varying $g$ can also be understood as changing the time window $\tau$ during which detectors are monitored. Indeed, the average number of photon pairs produced within this window is given by $\nu=\sinh^2 g+\sinh^2 \bar g=2\sinh^2 g$.

\begin{figure}
\includegraphics[width=0.5\textwidth]{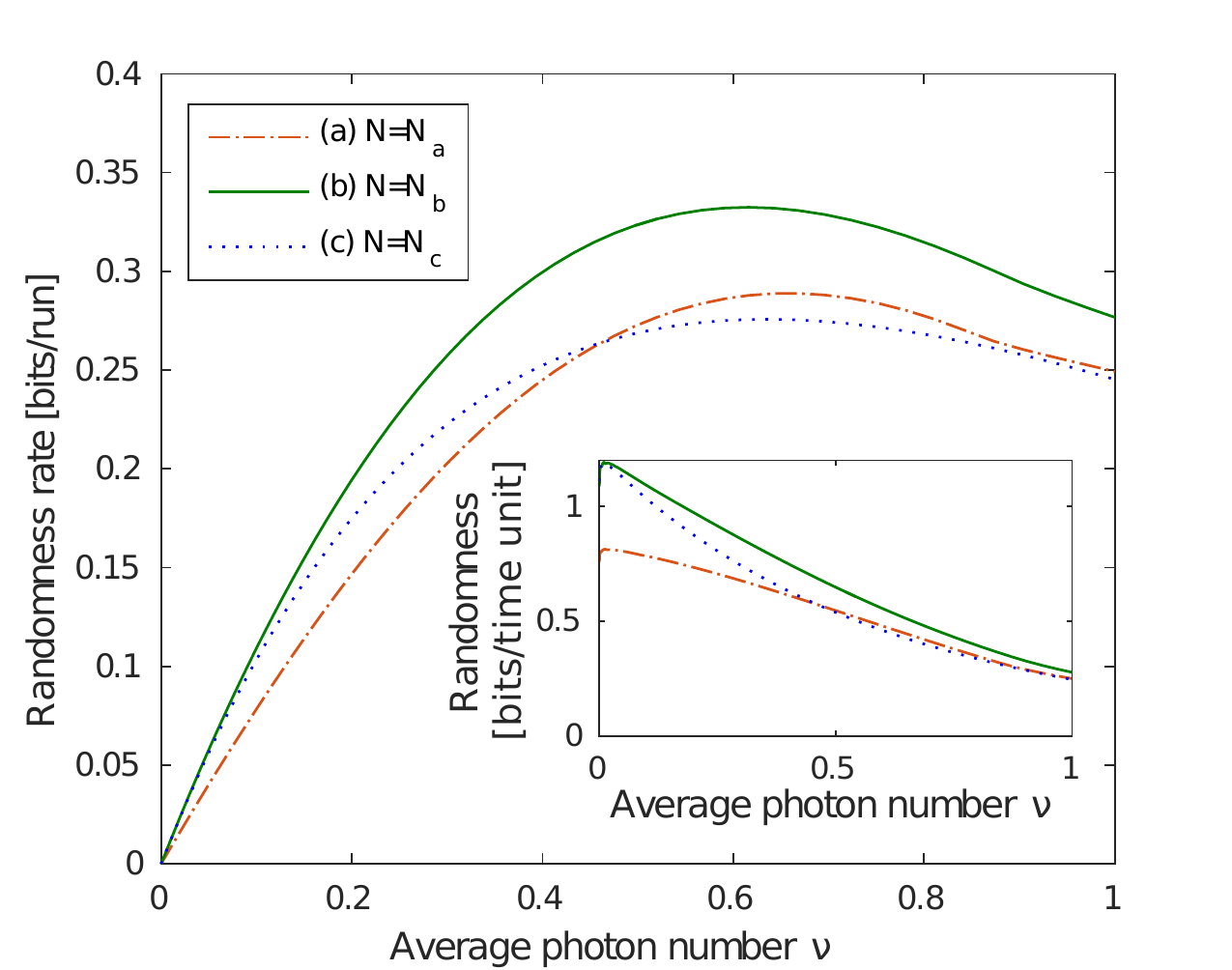}
\caption{\label{fig:squeezing} Randomness from an SPDC source when setting the polarization measurement according to the standard CHSH settings, as a function of the average number of photon pairs produced in each detection window. No losses and unit efficiency detectors are assumed. The qualitative shape of the curves can be understood as follows: for small $g$, the generated state contains mostly the vacuum; for large $g$, the source generates several pairs, which worsens the statistics~\cite{MSBCA15}. Strategies (a), (b) and (c) certify various amounts of randomness. Here and in the following figures, all the curves are normalised to the same number of runs, namely the total number of runs. Inset: Randomness certified in a given time period when the length of a time window varies (and the number of time windows varies accordingly). This curve is obtained at constant pumping $g$.}
\end{figure}

Figure~\ref{fig:squeezing} shows the randomness per run obtained when setting the polarization measurement according to the standard CHSH settings. The various discarding strategies yield different amounts of certified randomness, the largest amount being obtained using strategy (b).

One may be tempted to infer that, for randomness extraction, SPDC sources should be operated with detection window at $\nu\sim0.6$. However, this is the amount of randomness \textit{per run, not per time}. For a given pump power, decreasing the window size $\tau$ decreases the average number of photon pairs in a proportional manner: $\nu \propto \tau$. At the same time, the number of time windows increases as $\sim 1/\tau\propto 1/\nu$. If $f(\nu)$ denotes the randomness rate per time window, the randomness that can be certified in a given time interval is thus given, up to a constant factor, by $f(\nu)/\nu$. This quantity is plotted in the inset of Figure~\ref{fig:squeezing}, where one can see that total amount of randomness certified is larger when $\nu$ is small, i.e. the time window $\tau$ is small. Therefore, in the asymptotic limit of infinitely many runs, one should set $\tau\rightarrow0$ to get more randomness \emph{per time} against an i.i.d adversary. In this case, the observed data set is dominated by double no-detection events, which reinforces the relevance of our post-selection approach. The regime of small $\nu$ is also the regime in which optical experiments closing the detection loophole have been performed~\cite{Shalm15,Giustina15,christensen13}, for a different reason: in the presence of losses and imperfect detectors, the Bell violation disappears if too many pairs are created, while is preserved in the limit of small windows~\cite{MPRG02}.

\subsection{Imperfect detectors, small squeezing}\label{heraldedSource}

\begin{figure}
\includegraphics[width=0.5\textwidth]{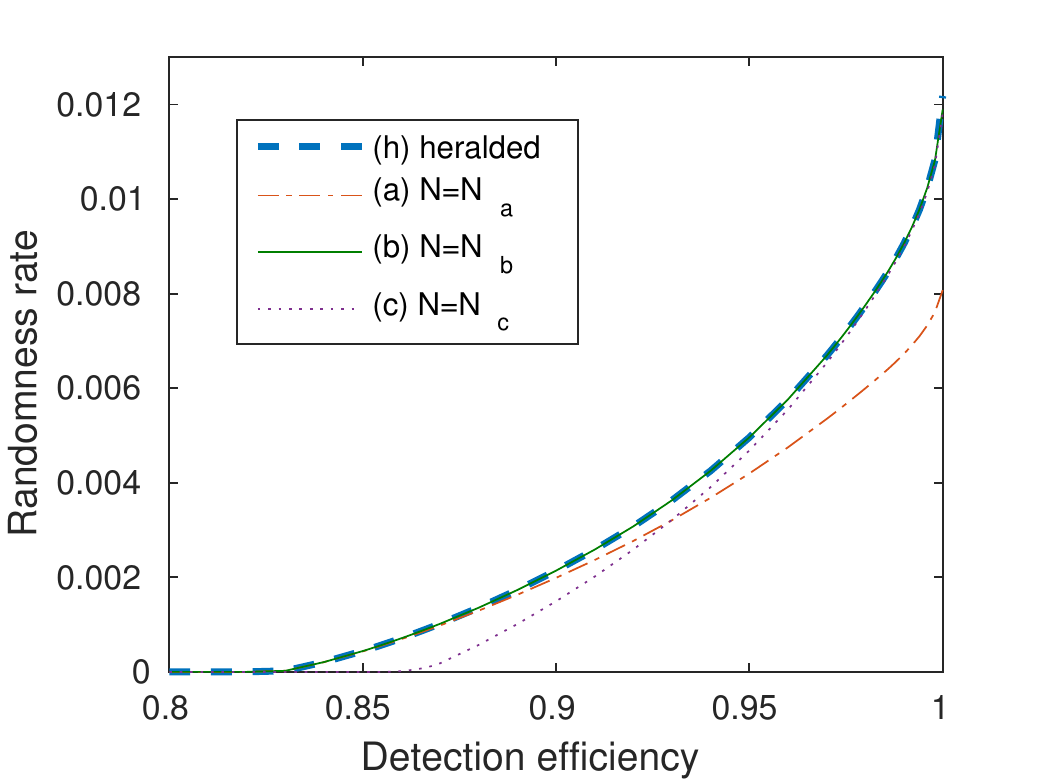}
\caption{\label{fig:ideal_source} Randomness from a singlet with finite detection efficiency. Curves (b) and (h) coincide almost perfectly and approach 0 at the detection loophole limit 0.828~\cite{mermin86}.}
\end{figure}

\begin{figure}
\includegraphics[width=0.5\textwidth]{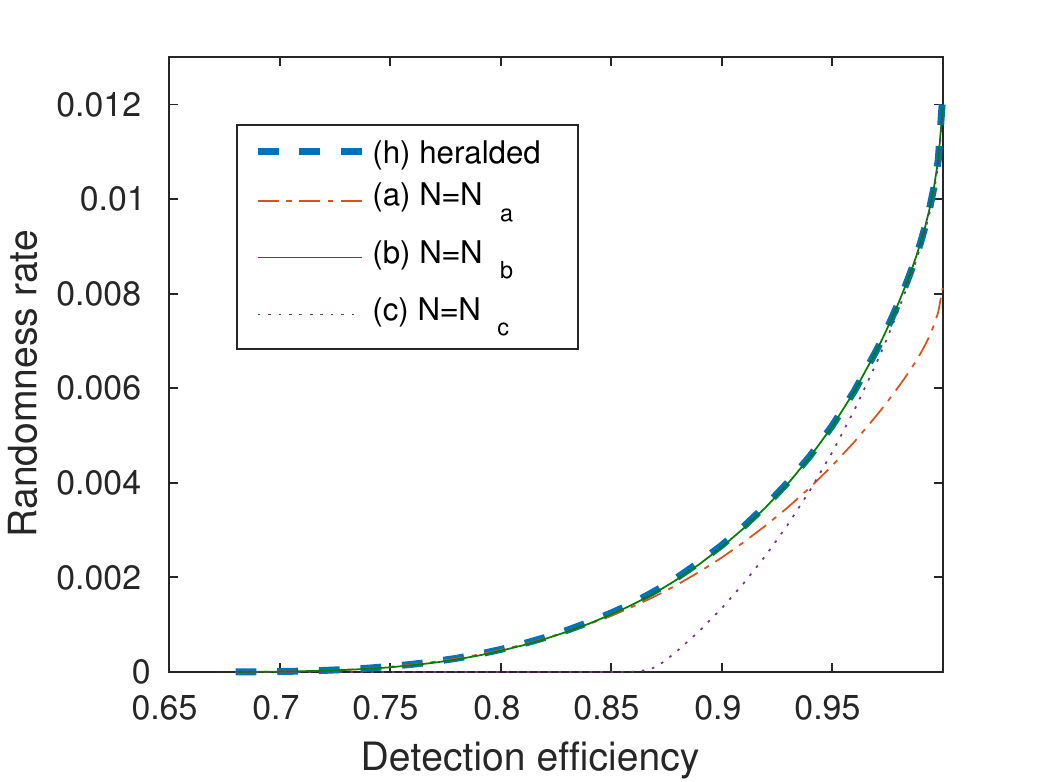}
\caption{\label{fig:Eberhard}  Randomness from Eberhard correlations. Curves (b) and (h) coincide and approach 0 at the Eberhard limit of 2/3~\cite{eberhard93}. Two recent experiments used this Eberhard correlations. In Ref.~\cite{Giustina15}, the overall efficiencies are estimated at 78.6\% for Alice and 76.2\% for Bob; in Ref.~\cite{Shalm15}, at 74.7\% for Alice and 75.6\% for Bob. Thus, strategies (a) and (b) would extract a very similar (small) amount of randomness. If efficiencies are increased in the future, strategy (b) should be preferred.}
\end{figure}

For the reasons just mentioned, we focus now on $g,\bar g<<1$. (i.e. small $\nu$). In this case, a large number of no-detection events is expected. In spite of this, we are going to see that strategy (b) continues to perform better than the others. Concretely, we choose to fix the average number of photon per detection window as $\nu=0.01$. The state produced by the source can be approximated to first order in $g$ and $\bar g$ by
\begin{equation}
\ket{\psi}\propto \ket{0} + (\tanh(g)a_H^\dag b_V^\dag - \tanh(\bar g) a_V^\dag b_H^\dag)\ket{0}\,.
\end{equation}
In analogy with the partially entangled state $\cos \theta \ket{01} - \sin\theta\ket{10}$, we define the entanglement parameter of the state as $\theta=\arctan(\tanh\bar g / \tanh g)$. 

We now introduce finite detection efficiency $\eta$ and study how the certification of randomness varies with this parameter. We then consider two families of correlations. In the first, the two-photon state is maximally entangled, i.e. with $\theta=\pi/4$, and we fix the standard CHSH polarization measurements. The expected randomness per run as a function of $\eta$ is shown in Figure~\ref{fig:ideal_source}. We note that no randomness can be extracted if $\eta\leq 82.8\%$ which is known to be the boundary at which those correlations can be explained with a local model exploiting the detection loophole. The second case is that of Eberhard's famous study~\cite{eberhard93}, in which the entanglement parameter $\theta$ depends on the detector efficiency $\eta$, and Alice's measurements are parametrized by two angles $\alpha_0,\alpha_1$ which also depend on $\eta$. These parameters are chosen to optimize the violation of a lifting~\cite{P05lifting} of the CHSH inequality, in the case where exactly one pair of photons is measured, for each value of $\eta$. The resulting randomness rate is plotted in Figure~\ref{fig:Eberhard}. Again, no randomness can be extracted below the known detection loophole threshold $\eta\leq 66.6\%$.

In both cases we notice again that, within a numerical precision $\sim10^{-5}$, strategy (b) certifies the largest amount of randomness and in fact recovers the result that one would obtained with a heralded source (h). The expected proportion of discarded events is $\sim (1-\nu)+\nu(1-\eta)^2$, which can be substantial: it is larger than 99\% in our case for all $\eta$. Strategy (c), i.e. removing all events where some no-detection occurred, results in clearly lower randomness per run; and for efficiencies lower than 86\% and 85\%, no randomness at all is even certified. This kind of post-selection is thus too strong if one is interested in certifying an optimal amount of randomness. Strategy (a) certifies essentially the maximum amount of randomness for efficiencies $\eta\lesssim 90\%$, but would become suboptimal as efficiency increases.



\section{Understanding why one certifies more randomness from a subset of data}\label{sec:perfectDet}

Let us stress again that in Figures \ref{fig:squeezing}-\ref{fig:Eberhard}, all the curves are normalised to the same number of runs, the total one. Thus, they show that \textit{if a suitable small fraction of the symbols is processed, a strictly larger amount of total randomness can be certified, as compared to the case where all the symbols are processed}. In order to shed light on this behavior, we consider a simplified model in which the source emits a perfect maximally-entangled state with probability $\nu$, and the vacuum otherwise (in other words, compared to the previous section, we neglect completely the possibility of double detections in each party's measurement setup). We also work at perfect detection efficiency $\eta=1$. The statistics observed with such a source can be written as
\begin{equation}
p(ab|xy) = \left\{ \begin{array}{ll}
          \nu \frac{1}{4}\left(1+(-1)^{a+b+xy}\frac{1}{\sqrt{2}}\right) & \mbox{if $a,b\in \{0,1\}$},\\
          (1-\nu) & \mbox{if $a=b=\varnothing$}.
\end{array} \right.\,.\label{eq:test_correlations}
\end{equation}
Notice that, for the source efficiency $\nu=\frac{1}{2}$, these correlations can be seen as the scrambled version of the two-day extreme situation mentioned in the introduction.

\begin{figure}
\includegraphics[width=0.5\textwidth]{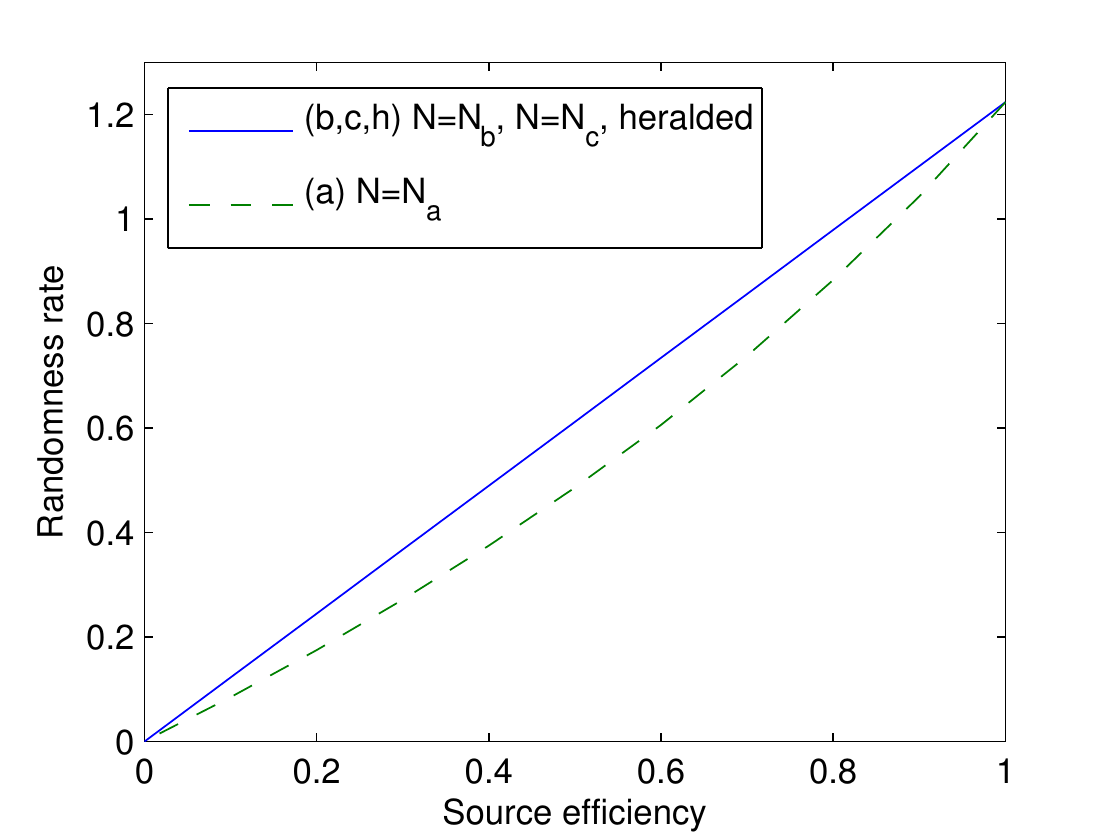}
\caption{\label{fig:ideal_detectors} Randomness from a singlet produced with finite probability $\nu$, with $\eta=1$. Curves (b) and (c) are identical, since there are no events with one detection and one no-detection in the raw data (the post-selection procedures (b) and (c) are actually the same for this correlation). Curve (h), which gives the randomness from raw string of outcomes upon the heralding of a successful preparation of the state (i.e. randomness from the correlation~\ref{eq:test_correlations}), exactly coincides with curves (b) and (c). Curve (a) lies below the other ones.}
\end{figure}


In Figure~\ref{fig:ideal_detectors}, we show how much randomness can be certified for these statistics when $\nu$ varies. In this case, the lower bound on the randomness computed from the raw data is this time consistantly \textit{lower} than the one obtained after removing double no-detections from the data. In fact, after discarding double no-detections, the same amount of randomness that could be certified if the source was heralded is recovered (i.e. it is proportional to the source efficiency $\nu$).

We thus recover the same behaviour as discussed in Section~\ref{heraldedSource} and in the two-day example of the introduction. If we don't consider it an overwhelmingly improbable fluctuation, the two-day example clearly suggests a non-i.i.d. process, for which the possibility of identifying two separate processes is easy to understand. Here, on the contrary, the statistics are manifestly i.i.d. --- and nevertheless, the extraction of randomness based on the single-run frequencies $p(ab|xy)$ can be improved. We are going to show that the cause is the same: \textit{because of the structure of the correlations, one can actually identify the presence of two distinct processes, and the post-selection of detection events happens to capture this fact}. That the alternation between the two processes is done in an i.i.d. way, instead of a disruptive way as in the two-day example, eventually does not matter.

Note first that by definition $p(ab|xy)$ has the block structure $p(ab|xy)=\nu\, q(ab|xy)+(1-\nu)\, r(ab|xy)$ where
\begin{equation}
q(ab|xy) =  \begin{array}{cc||ccc|ccc}
 & \mathbf{y} & & \mathbf{0} & & & \mathbf{1}&\\
\mathbf{x}&\it{a \backslash b} &\it{0}&\it{1}&\it{\varnothing} &\it{0}&\it{1}&\it{\varnothing}\\ \hline\hline
& \it{0}& q_1 & q_2 & 0  & q_5 & q_6 & 0\\
\mathbf{0}&\it{1}& q_3&  q_4& 0 & q_7&q_8 & 0\\
&\it{\varnothing} & 0 & 0 & 0 & 0 & 0 & 0\\ \hline
& \it{0}& q_9&  q_{10}& 0  &q_{13}  & q_{14}& 0\\
\mathbf{1}&\it{1}&q_{11} &  q_{12}& 0 & q_{15}& q_{16}& 0\\
&\it{\varnothing} & 0 & 0 & 0 & 0 & 0 & 0
\end{array}
\end{equation}
and
\begin{equation}
r(ab|xy) =  \begin{array}{cc||ccc|ccc}
 & \mathbf{y} & & \mathbf{0} & & & \mathbf{1}&\\
\mathbf{x}&\it{a \backslash b} &\it{0}&\it{1}&\it{\varnothing} &\it{0}&\it{1}&\it{\varnothing}\\ \hline\hline
& \it{0}& 0& 0 & 0  & 0 & 0 & 0\\
\mathbf{0}&\it{1}& 0&  0&0 & 0&0 & 0\\
&\it{\varnothing} & 0 & 0 & 1 & 0 & 0 & 1\\ \hline
& \it{0}& 0&  0& 0  &0  & 0& 0\\
\mathbf{1}&\it{1}&0 &  0& 0 & 0& 0& 0\\
&\it{\varnothing} & 0 & 0 & 1 & 0 & 0 & 1
\end{array}
\end{equation}
It follows that in the decomposition $p(ab|xy)=\sum_{a'b'\in\mathcal{V}}  \tilde p_{a'b'}(ab|xy)$ in the second line of the program (\ref{goptim}), every $\tilde p_{a'b'}(ab|xy)$ must also have this block structure, since if $p(ab|xy)$ is equal to zero for some $a,b,x,y$ then $\tilde p_{a'b'}(ab|xy)$ must also necessarily be equal to zero. We can thus write $\tilde p_{a'b'}(ab|xy)=\nu_{a'b'}\,q_{a'b'}(ab|xy)+(1-\nu_{a'b'})\,r(ab|xy)$ where $q_{a'b'}(ab|xy)$ is normalized and has the same general form as $q(ab|xy)$ above. The condition $p(ab|xy)=\sum_{a'b'\in\mathcal{V}}  \tilde p_{a'b'}(ab|xy)$ is then equivalent to $\sum_{a'b'\in\mathcal{V}}\nu_{a'b'}=\nu$ and $\sum_{a'b'\in\mathcal{V}} q_{a'b'}(ab|xy)=\nu q(ab|xy)$. 

Furthermore, when we post-select events according to (b) or (c), the effective set of valid symbols is in both cases $\mathcal{V}=\{00,01,10,11\}$ since outcome pairs $0\varnothing$, $1\varnothing$, $\varnothing 0$, $\varnothing 1$ have zero probability. The objective value in (\ref{goptim}) therefore only involves the $q_{a'b'}(ab|xy)$ part and is equal to $1/\nu\, \max \sum_{ab\in\mathcal{V}} \nu_{ab}\,q_{ab}(ab|xy)$, where we used that $p_{\bar x\bar y}=\sum_{ab\in\mathcal{V}} \nu_{ab}=\nu$. 

All together, we can thus rewrite the optimization (\ref{goptim}) as
\begin{eqnarray}
G_{\bar x \bar y} = \frac{1}{\nu} &\max & \sum_{ab\in\mathcal{V}} \nu_{ab} q_{ab}(ab|\bar x\bar y)\nonumber\\
& s.t. & \sum_{a'b'\in\mathcal{V}} \nu_{a'b'}\, q_{a'b'}(ab|xy)=\nu \,q(ab|xy)\\
& & q_{a'b'}(ab|xy) \in Q\nonumber
\end{eqnarray}
where $\mathcal{Q}$ denotes the set of normalized quantum correlations. Defining $\tilde q_{a'b'}(ab|xy)=\nu_{a'b'}/\nu\	 \times q_{ab}(ab|\bar x\bar y)$, we can further rewrite it as 
\begin{eqnarray}
G_{\bar x \bar y} = &\max & \sum_{ab\in\mathcal{V}} \tilde q_{ab}(ab|\bar x\bar y)\nonumber\\
& s.t. & \sum_{a'b'\in\mathcal{V}} \tilde q_{a'b'}(ab|xy)=q(ab|xy)\\
& & \tilde q_{a'b'}(ab|xy) \in \tilde Q\nonumber
\end{eqnarray}
This optimization is nothing but the one associated to a heralded source characterized by the correlations $q(ab|xy)$ and explains why curve (h) of Figure~\ref{fig:ideal_detectors} coincides with curves (b) and (c).

%
%
%

\section{Going beyond i.i.d. for the source}\label{sec:noniid}

In this section, we are going to relaxing the i.i.d. assumption for the source. We won't be able to derive bounds for the extraction of randomness from the most general non-i.i.d. source. But we are going to provide two example of non-i.i.d. strategies that are strictly more powerful than i.i.d. strategies \textit{even in the asymptotic limit of infinitely many runs}. To our knowledge, this is a feature not found in previous works on randomness from Bell tests~\cite{PAM10,BCHKP02,gill14} or on quantum key distribution~\cite{rennerThesis}. In the strategies we found, the adversary exploits the knowledge of whether each outcome is kept or discarded. As mentioned in Section~\ref{sec:method}, it would be definitely reasonable not to reveal anything, but such scenario may introduce other security concerns (e.g. the raw key is private conditional on some other information being kept private). 

Specifically, suppose that the outcomes of run $k$ are valid, i.e. they are kept for the raw key; the adversary would like to know their value. In a non-i.i.d. case, the fact of keeping or discarding the outcome at run $k+1$, an information which we assume the adversary will learn, may leak some information about the outcome that is kept at run $k$. This is similar to the argument of~\cite{BCK13} against reusing QKD devices in the device-independent level of characterization~\cite{LTBS14}. Notice that this behaviour does not require the adversary to have tampered with the device in a malicious way, it may be simply a defect of fabrication that the adversary is aware of. For instance, suppose that the detector corresponding to outcome 0 has an inordinately long jitter time compared to the other detector: if a detection happens at run $k+1$, it means that the outcome at run $k$ was 1; if no detection, the outcome at run $k$ was most probably 0.

\subsection{First example}

The simplest example we found requires both Alice's and Bob's devices to depend on the previous inputs and outputs of \textit{both sides}. Note that this is not in contradiction with the basic assumption in all device-independent protocols that the two boxes are non-communicating, since this assumption must only be verified during the measurement runs. Between measurement runs, however, boxes could in principle be free to communicate. For instance, before the measurement runs, the boxes may open a door within a small time interval to let enter incoming quantum systems, those generated by and coming from the source. Malicious boxes could take advantage of this interval to exchange the inputs and outputs obtained in previous runs. In the next subsection, we will present a more convoluted example that does not require signalling between the boxes, and thus which also works if measure are taken to insure that the boxes do not exchange such kind of information between measurement runs.

Consider the i.i.d. correlations obtained when the parties measure a singlet with probability $\nu$, and nothing with probability $1-\nu$. We have encountered this situation in paragraph \ref{sec:perfectDet}: for any $\nu>0$, some randomness remains in the non-discarded outcomes (see Figure~\ref{fig:ideal_detectors}).

In all existing protocols, the amount of randomness that is extracted is determined from a statistical test which is based on the input and output pair counts $\#(x,y)$ and $\#(a,b)$ (or simply relative outcome frequencies $\#(a,b)/\#(x,y)$. However, the same statistics obtained for $\nu=2/5$ can be obtained with high probability when measurements are always performed on a perfect singlet, but runs with double no-detections are artificially added by using the following non-i.i.d. rule:
\begin{equation}
\begin{array}{c|cccc}
\text{singlet outcomes }(a,b) & \multicolumn{3}{c}{\text{following runs}}\\
\hline
(0,0) & M& & &\\
(0,1) & (\varnothing,\varnothing) & M& &\\
(1,0) & (\varnothing,\varnothing) & (\varnothing,\varnothing) &M &\\
(1,1) & (\varnothing,\varnothing) & (\varnothing,\varnothing) & (\varnothing,\varnothing)& M\\
\end{array}
\end{equation}
where $M$ means that an usual measurement is performed on the perfect singlet to determine the outcome of that run. 
In this case, counting the number of successive discarded events fully informs about the value of both parties' outcomes. Thus, in the non-i.i.d case, and allowing signalling from one box to the other between measurement runs, no private randomness can be certified from a non heralded source characterized by $\nu\leq 2/5$ (unless some more complicated processing beyond looking at simple outcome counts is done).

\subsection{Second example}

The second example was found numerically. It is admittedly hard to find a narrative justification for it, besides the general intuition given above; but we describe it in detail since, to our knowledge, it is the first example in which a non-i.i.d. strategy actually outperforms the i.i.d. ones in a Bell scenario in the asymptotic limit.

\textit{Resources.} In each run, Alice and Bob share two binary variables $\lambda,\mu\in\{0,1\}$ and one out of five quantum correlations that we denote by $P_{j}$ with $j\in\{1,2,3\}$ and $P'_{\lambda}$. These correlations are such that Alice's box has three outcomes $\{0,1,\varnothing\}$, while Bob's box has only the two outcomes $\{0,1\}$: in other words, information about previous outcomes will be leaked out by Alice's box detection or no-detection events. We can write these correlations as above in the form of Collins-Gisin tables~\cite{CG04}:
\ba
P =  \begin{array}{cc||cc|cc}
 & \mathbf{ y} & & \mathbf{0} &  & \mathbf{1}\\
\mathbf{ x}&\it{ a \backslash b} &\it{0}&\it{1} &\it{0}&\it{1}\\ \hline\hline
& \it{0}& \cdot & \cdot   & \cdot & \cdot \\
\mathbf{0}&\it{1}& \cdot&  \cdot & \cdot&\cdot \\
&\it{\varnothing} & \cdot&  \cdot & \cdot&\cdot \\ \hline
& \it{0}& \cdot&  \cdot  &\cdot  & \cdot\\
\mathbf{1}&\it{1}&\cdot &  \cdot & \cdot& \cdot\\
&\it{\varnothing} &\cdot&  \cdot & \cdot&\cdot \end{array} 
&\equiv&
\begin{array}{c|c|c}
 1 & P_A(0|0)   & P_A(0|1)\\
\hline
    P_B(0|0)   & P(00|00)   & P(00|01)\\
    P_B(1|0)   & P(10|00)   & P(10|01)\\
\hline
    P_B(0|1)   & P(00|10)   & P(00|11)\\
    P_B(1|1)   & P(10|10)   & P(10|11)\\
\end{array}\nonumber\ea
because by no-signaling it holds $P(a1|xy)=P_A(a|x)-P(a0|xy)$ and $P(\varnothing b|xy)= P_B(b|y)- P(0 b|xy) -P(1 b|xy)$; and of course $\sum_aP(a|x)=\sum_bP(b|y)=1$. The example that we find uses:
\begin{align}
P_1 &= \begin{array}{c|c|c}
    1 & 0.4453   & 0.3121\\
\hline
    0.6570   & 0.1708   & 0.0394\\
         0       &  0        & 0\\
\hline
    0.3244   & 0.0247   & 0.2843\\
    0.4942   & 0.4195   & 0.0277\\
\end{array}\ ,\\ 
P_2 &= \begin{array}{c|c|c}
    1   & 0.8544   & 0.7373\\
\hline
         0        & 0       &  0\\
    0.8919   & 0.8381 &   0.7209\\
\hline
    0.2619   & 0.1165   & 0.2617\\
    0.4973   & 0.4972   & 0.2354\\
\end{array}\ ,\\ 
P_3 &= \begin{array}{c|c|c}
    1   & 0.6042   & 0.5429\\
\hline
    0.3979   & 0.0886   & 0.0365\\
    0.6021   & 0.5156   & 0.5064\\
\hline
    0.4588   & 0.1078   & 0.4267\\
    0.5412   & 0.4964   & 0.1162\\
\end{array}\,\\
P'_{\lambda=0} &= \begin{array}{c|c|c}
    1   & 0.6663   & 0.2038\\
\hline
    1   & 0.6663   & 0.2038\\
0 & 0 & 0\\
\hline
    0.2936   & 0.1393   & 0.1112\\
    0.7064   & 0.5270   & 0.0926\\
\end{array}\ ,\\ 
P'_{\lambda=1} &= \begin{array}{c|c|c}
    1   & 0.9996   & 0.0015\\
\hline
0&0&0\\
    1   & 0.9996   & 0.0015\\
\hline
    0.0010   & 0.0006   & 0.0004\\
    0.9990   & 0.9990   & 0.0011\\
\end{array}\ .
\end{align}

\textit{Protocol.} One starts with one of the three $P_j$'s. As long as $j=1$ or $j=2$, the next round will also use one of the three $P_j$'s. When $P_3$ was chosen, the next box will be $P'_{\lambda}$ with the value of $\lambda$ available in that run. Besides, if Alice's outcome from $P_3$ was $a=\mu$, in the next run Alice uses the box $P'_{\lambda}$; if the outcome was $a=1-\mu$, in the next run Alice ignores $P'_{\lambda}$ and outputs $\varnothing$. After this, the process starts again by selecting one of the three $P_j$'s. 

Now, 
when $x=0$, either outcome 0 or outcome 1 cannot occur for each potential correlation except $P_3$; and when $P_3$ is used, its outcomes is fully leaked out in the next run by the information of whether the subsequent outcome is kept or not, since $P'_{\lambda}(\varnothing|x)=0$.

One can check, however, that it would not be possible to fully guess Alice's outcome if the same outcome relative frequencies as the one generated by the above process where produced by devices behaving in an i.i.d manner. For instance, let us specify $q_1 = 0.4097$, $q_2 = 0.4992$, $q_3 = 0.0911$ as the frequencies at which the $P_j$'s are chosen; and $p(\lambda=0) = 1 - p(\lambda=1)=0.0013$, $p(\mu=0) = p(\mu=1) = 1/2$. The expected relative frequencies in the asymptotic limit are then peaked around the following values
\begin{widetext}
\begin{align}
P = \frac{q_1P_1 + q_2P_2 + q_3P_3 + q_3(p(\lambda=0) (P'_0 + {P'_0}^B)/2 + p(\lambda=1) (P'_1 + {P'_1}^B)/2)}{q_1+q_2+2q_3}
=\begin{array}{c|c|c}
    1   & 0.6919   & 0.5000\\
\hline
    0.2800   & 0.0716   & 0.0178\\
    0.5000   & 0.4681   & 0.3722\\
\hline
    0.2800   & 0.0716   & 0.2621\\
    0.5000   & 0.4681   & 0.1279\\
\end{array}\ ,
\end{align}
\end{widetext}
where ${P'_\lambda}^B$ denote the correlations obtained when Bob uses $P'_\lambda$ and Alice outputs $\varnothing$. Applying our i.i.d. programme to these correlations, one can show that in case Alice uses $x=0$ and the run is not discarded, the guessing probability on her outcome is upper-bounded by 0.9874.

\section{Conclusion}

This work stems from the general remark that randomness extraction does not need to be performed on all of the raw data and can be done by blocks, or on a subset of data. In the context of randomness certification by Bell inequalities, we have investigated in a simple scenario whether this could provide an advantage when post-selecting detection events, which is relevant for photonics Bell tests. Because we estimate the randomness present in a subset of data \textit{conditioned} on the knowledge of the whole set of data, this certification does not open the detection loophole.

Naively, one could a priori think that ``full detection" events, where a detection happens on both side, are the most important for randomness certification and that discarding all other events would influence only negligibly the randomness rate. However, our findings show for several physically-motivated models of the observed statistics that this is not the case. In particular, Figure~\ref{fig:ideal_source} and Figure~\ref{fig:Eberhard} show that the resistance to detection inefficiencies is substantially lower (up to 20\% for the scenario Figure~\ref{fig:Eberhard}) when the post-selected data does not contain any occurrence of a no-detection event. 

The physical intuition that the double no-detection events contain almost no randomness is, however, vindicated. In some cases, the post-selection actually help identify a better way of reading the data. From a practical perspective, our work suggests the possibility of hashing a small post-selected subset of the original data, thereby reducing the needed seed length, and ultimately the computational time. However, one should still embed this idea within a full randomness certification protocol, in particular one that can deal with finite statistics and non-i.i.d. devices. 

Regarding this last point, the physical intuition that double no-detection events can safely be discarded, as vindicated by our numerical results in an i.i.d. setting, should, however, be contrasted with the example of Section~\ref{sec:noniid} in which we prove that non-i.i.d. strategies outperform i.i.d. ones even in the asymptotic limit of infinitely many runs, something that had not been reported previously in the context of Bell inequalities. Whether these strategies are actually harmful in a more general and realistic case remains to be determined. 

In particular, we remind that for simplicity we have performed our analysis assuming that the adversary gets to know which runs are kept and which ones are discarded in the post-selection. This scenario is rather artificial for randomness generation, insofar as the two boxes for the Bell experiment don't need to be in separate labs. Relaxing this assumption could increase the randomness rate and the security of the final string. Specifically, the non-i.i.d. attacks of Section~\ref{sec:noniid} would not apply anymore in this case.

\begin{acknowledgments}
We thank Nicolas Brunner and Nicolas Sangouard for stimulating discussions.

This work is funded by the Singapore Ministry of Education (partly through the Academic Research Fund Tier 3 MOE2012-T3-1-009) and by the National Research Foundation of Singapore. GdlT acknowledges support from Spanish FPI grant (FIS2010-14830) and the subsequent hospitality from CQT. S.P. acknowledges financial support from the European Union under the project QALGO, from the F.R.S.-FNRS under the project DIQIP, and by the Interuniversity Attraction Poles program of the Belgian Science Policy Office under the grant IAP P7-35 photonics@be. S. P. is a Research Associate of the Fonds de la Recherche Scientifique F.R.S.-FNRS (Belgium).

\end{acknowledgments}

\bibliography{citations}

\begin{thebibliography}{36}%
\makeatletter
\providecommand \@ifxundefined [1]{%
 \@ifx{#1\undefined}
}%
\providecommand \@ifnum [1]{%
 \ifnum #1\expandafter \@firstoftwo
 \else \expandafter \@secondoftwo
 \fi
}%
\providecommand \@ifx [1]{%
 \ifx #1\expandafter \@firstoftwo
 \else \expandafter \@secondoftwo
 \fi
}%
\providecommand \natexlab [1]{#1}%
\providecommand \enquote  [1]{``#1''}%
\providecommand \bibnamefont  [1]{#1}%
\providecommand \bibfnamefont [1]{#1}%
\providecommand \citenamefont [1]{#1}%
\providecommand \href@noop [0]{\@secondoftwo}%
\providecommand \href [0]{\begingroup \@sanitize@url \@href}%
\providecommand \@href[1]{\@@startlink{#1}\@@href}%
\providecommand \@@href[1]{\endgroup#1\@@endlink}%
\providecommand \@sanitize@url [0]{\catcode `\\12\catcode `\$12\catcode
  `\&12\catcode `\#12\catcode `\^12\catcode `\_12\catcode `\%12\relax}%
\providecommand \@@startlink[1]{}%
\providecommand \@@endlink[0]{}%
\providecommand \url  [0]{\begingroup\@sanitize@url \@url }%
\providecommand \@url [1]{\endgroup\@href {#1}{\urlprefix }}%
\providecommand \urlprefix  [0]{URL }%
\providecommand \Eprint [0]{\href }%
\providecommand \doibase [0]{http://dx.doi.org/}%
\providecommand \selectlanguage [0]{\@gobble}%
\providecommand \bibinfo  [0]{\@secondoftwo}%
\providecommand \bibfield  [0]{\@secondoftwo}%
\providecommand \translation [1]{[#1]}%
\providecommand \BibitemOpen [0]{}%
\providecommand \bibitemStop [0]{}%
\providecommand \bibitemNoStop [0]{.\EOS\space}%
\providecommand \EOS [0]{\spacefactor3000\relax}%
\providecommand \BibitemShut  [1]{\csname bibitem#1\endcsname}%
\let\auto@bib@innerbib\@empty
\bibitem [{\citenamefont {Metropolis}\ and\ \citenamefont {Ulam}(1949)}]{MU49}%
  \BibitemOpen
  \bibfield  {author} {\bibinfo {author} {\bibfnamefont {N.}~\bibnamefont
  {Metropolis}}\ and\ \bibinfo {author} {\bibfnamefont {S.}~\bibnamefont
  {Ulam}},\ }\href {http://www.jstor.org/stable/2280232} {\bibfield  {journal}
  {\bibinfo  {journal} {Journal of the American Statistical Association}\
  }\textbf {\bibinfo {volume} {44}},\ \bibinfo {pages} {pp. 335} (\bibinfo
  {year} {1949})}\BibitemShut {NoStop}%
\bibitem [{\citenamefont {Motwani}\ and\ \citenamefont
  {Raghavan}(1995)}]{MR95}%
  \BibitemOpen
  \bibfield  {author} {\bibinfo {author} {\bibfnamefont {R.}~\bibnamefont
  {Motwani}}\ and\ \bibinfo {author} {\bibfnamefont {P.}~\bibnamefont
  {Raghavan}},\ }\href@noop {} {\emph {\bibinfo {title} {Randomized
  Algorithms}}}\ (\bibinfo  {publisher} {Cambridge University Press},\ \bibinfo
  {address} {New York, NY, USA},\ \bibinfo {year} {1995})\BibitemShut {NoStop}%
\bibitem [{\citenamefont {Vadhan}(2011)}]{vadhan11}%
  \BibitemOpen
  \bibfield  {author} {\bibinfo {author} {\bibfnamefont {S.~P.}\ \bibnamefont
  {Vadhan}},\ }\href {\doibase 10.1561/0400000010} {\bibfield  {journal}
  {\bibinfo  {journal} {Foundations and Trends® in Theoretical Computer
  Science}\ }\textbf {\bibinfo {volume} {7}},\ \bibinfo {pages} {1} (\bibinfo
  {year} {2011})}\BibitemShut {NoStop}%
\bibitem [{\citenamefont {{Colbeck}}(2009)}]{colbeckThesis}%
  \BibitemOpen
  \bibfield  {author} {\bibinfo {author} {\bibfnamefont {R.}~\bibnamefont
  {{Colbeck}}},\ }\emph {\bibinfo {title} {{Quantum And Relativistic Protocols
  For Secure Multi-Party Computation}}},\ \href@noop {} {Ph.D. thesis},\
  \bibinfo  {school} {PhD Thesis, 2009} (\bibinfo {year} {2009})\BibitemShut
  {NoStop}%
\bibitem [{\citenamefont {Pironio}\ \emph
  {et~al.}(2010{\natexlab{a}})\citenamefont {Pironio}, \citenamefont
  {Ac\'{\i}n}, \citenamefont {Massar}, \citenamefont {de~la Giroday},
  \citenamefont {Matsukevich}, \citenamefont {Maunz}, \citenamefont
  {Olmschenk}, \citenamefont {Hayes}, \citenamefont {Luo}, \citenamefont
  {Manning},\ and\ \citenamefont {Monroe}}]{PAM10}%
  \BibitemOpen
  \bibfield  {author} {\bibinfo {author} {\bibfnamefont {S.}~\bibnamefont
  {Pironio}}, \bibinfo {author} {\bibfnamefont {A.}~\bibnamefont {Ac\'{\i}n}},
  \bibinfo {author} {\bibfnamefont {S.}~\bibnamefont {Massar}}, \bibinfo
  {author} {\bibfnamefont {A.~B.}\ \bibnamefont {de~la Giroday}}, \bibinfo
  {author} {\bibfnamefont {D.~N.}\ \bibnamefont {Matsukevich}}, \bibinfo
  {author} {\bibfnamefont {P.}~\bibnamefont {Maunz}}, \bibinfo {author}
  {\bibfnamefont {S.}~\bibnamefont {Olmschenk}}, \bibinfo {author}
  {\bibfnamefont {D.}~\bibnamefont {Hayes}}, \bibinfo {author} {\bibfnamefont
  {L.}~\bibnamefont {Luo}}, \bibinfo {author} {\bibfnamefont {T.~A.}\
  \bibnamefont {Manning}}, \ and\ \bibinfo {author} {\bibfnamefont
  {C.}~\bibnamefont {Monroe}},\ }\href@noop {} {\bibfield  {journal} {\bibinfo
  {journal} {Nature}\ }\textbf {\bibinfo {volume} {464}},\ \bibinfo {pages}
  {1021} (\bibinfo {year} {2010}{\natexlab{a}})}\BibitemShut {NoStop}%
\bibitem [{\citenamefont {Pironio}\ and\ \citenamefont {Massar}(2013)}]{PM11}%
  \BibitemOpen
  \bibfield  {author} {\bibinfo {author} {\bibfnamefont {S.}~\bibnamefont
  {Pironio}}\ and\ \bibinfo {author} {\bibfnamefont {S.}~\bibnamefont
  {Massar}},\ }\href {\doibase 10.1103/PhysRevA.87.012336} {\bibfield
  {journal} {\bibinfo  {journal} {Phys. Rev. A}\ }\textbf {\bibinfo {volume}
  {87}},\ \bibinfo {pages} {012336} (\bibinfo {year} {2013})}\BibitemShut
  {NoStop}%
\bibitem [{\citenamefont {Vazirani}\ and\ \citenamefont {Vidick}(2012)}]{VV12}%
  \BibitemOpen
  \bibfield  {author} {\bibinfo {author} {\bibfnamefont {U.}~\bibnamefont
  {Vazirani}}\ and\ \bibinfo {author} {\bibfnamefont {T.}~\bibnamefont
  {Vidick}},\ }\href@noop {} {\  (\bibinfo {year} {2012})},\ \bibinfo {note}
  {arXiv:1111.6054}\BibitemShut {NoStop}%
\bibitem [{\citenamefont {Bancal}\ \emph {et~al.}(2014)\citenamefont {Bancal},
  \citenamefont {Sheridan},\ and\ \citenamefont {Scarani}}]{bancal13more}%
  \BibitemOpen
  \bibfield  {author} {\bibinfo {author} {\bibfnamefont {J.-D.}\ \bibnamefont
  {Bancal}}, \bibinfo {author} {\bibfnamefont {L.}~\bibnamefont {Sheridan}}, \
  and\ \bibinfo {author} {\bibfnamefont {V.}~\bibnamefont {Scarani}},\ }\href
  {http://stacks.iop.org/1367-2630/16/i=3/a=033011} {\bibfield  {journal}
  {\bibinfo  {journal} {New Journal of Physics}\ }\textbf {\bibinfo {volume}
  {16}},\ \bibinfo {pages} {033011} (\bibinfo {year} {2014})}\BibitemShut
  {NoStop}%
\bibitem [{\citenamefont {Nieto-Silleras}\ \emph {et~al.}(2014)\citenamefont
  {Nieto-Silleras}, \citenamefont {Pironio},\ and\ \citenamefont
  {Silman}}]{silleras14more}%
  \BibitemOpen
  \bibfield  {author} {\bibinfo {author} {\bibfnamefont {O.}~\bibnamefont
  {Nieto-Silleras}}, \bibinfo {author} {\bibfnamefont {S.}~\bibnamefont
  {Pironio}}, \ and\ \bibinfo {author} {\bibfnamefont {J.}~\bibnamefont
  {Silman}},\ }\href {http://stacks.iop.org/1367-2630/16/i=1/a=013035}
  {\bibfield  {journal} {\bibinfo  {journal} {New Journal of Physics}\ }\textbf
  {\bibinfo {volume} {16}},\ \bibinfo {pages} {013035} (\bibinfo {year}
  {2014})}\BibitemShut {NoStop}%
\bibitem [{Note1()}]{Note1}%
  \BibitemOpen
  \bibinfo {note} {Notice that the difference grows linearly with $N$, thus it
  cannot be accounted for by finite-size corrections related to processing two
  $N$-symbol sets instead of a single $2N$-symbol one.}\BibitemShut {Stop}%
\bibitem [{\citenamefont {Christensen}\ \emph {et~al.}(2013)\citenamefont
  {Christensen}, \citenamefont {McCusker}, \citenamefont {Altepeter},
  \citenamefont {Calkins}, \citenamefont {Gerrits}, \citenamefont {Lita},
  \citenamefont {Miller}, \citenamefont {Shalm}, \citenamefont {Zhang},
  \citenamefont {Nam}, \citenamefont {Brunner}, \citenamefont {Lim},
  \citenamefont {Gisin},\ and\ \citenamefont {Kwiat}}]{christensen13}%
  \BibitemOpen
  \bibfield  {author} {\bibinfo {author} {\bibfnamefont {B.~G.}\ \bibnamefont
  {Christensen}}, \bibinfo {author} {\bibfnamefont {K.~T.}\ \bibnamefont
  {McCusker}}, \bibinfo {author} {\bibfnamefont {J.~B.}\ \bibnamefont
  {Altepeter}}, \bibinfo {author} {\bibfnamefont {B.}~\bibnamefont {Calkins}},
  \bibinfo {author} {\bibfnamefont {T.}~\bibnamefont {Gerrits}}, \bibinfo
  {author} {\bibfnamefont {A.~E.}\ \bibnamefont {Lita}}, \bibinfo {author}
  {\bibfnamefont {A.}~\bibnamefont {Miller}}, \bibinfo {author} {\bibfnamefont
  {L.~K.}\ \bibnamefont {Shalm}}, \bibinfo {author} {\bibfnamefont
  {Y.}~\bibnamefont {Zhang}}, \bibinfo {author} {\bibfnamefont {S.~W.}\
  \bibnamefont {Nam}}, \bibinfo {author} {\bibfnamefont {N.}~\bibnamefont
  {Brunner}}, \bibinfo {author} {\bibfnamefont {C.~C.~W.}\ \bibnamefont {Lim}},
  \bibinfo {author} {\bibfnamefont {N.}~\bibnamefont {Gisin}}, \ and\ \bibinfo
  {author} {\bibfnamefont {P.~G.}\ \bibnamefont {Kwiat}},\ }\href {\doibase
  10.1103/PhysRevLett.111.130406} {\bibfield  {journal} {\bibinfo  {journal}
  {Phys. Rev. Lett.}\ }\textbf {\bibinfo {volume} {111}},\ \bibinfo {pages}
  {130406} (\bibinfo {year} {2013})}\BibitemShut {NoStop}%
\bibitem [{\citenamefont {Eberhard}(1993)}]{eberhard93}%
  \BibitemOpen
  \bibfield  {author} {\bibinfo {author} {\bibfnamefont {P.~H.}\ \bibnamefont
  {Eberhard}},\ }\href {\doibase 10.1103/PhysRevA.47.R747} {\bibfield
  {journal} {\bibinfo  {journal} {Phys. Rev. A}\ }\textbf {\bibinfo {volume}
  {47}},\ \bibinfo {pages} {R747} (\bibinfo {year} {1993})}\BibitemShut
  {NoStop}%
\bibitem [{\citenamefont {Mermin}(1986)}]{mermin86}%
  \BibitemOpen
  \bibfield  {author} {\bibinfo {author} {\bibfnamefont {N.~D.}\ \bibnamefont
  {Mermin}},\ }\href {\doibase 10.1111/j.1749-6632.1986.tb12444.x} {\bibfield
  {journal} {\bibinfo  {journal} {Annals of the New York Academy of Sciences}\
  }\textbf {\bibinfo {volume} {480}},\ \bibinfo {pages} {422} (\bibinfo {year}
  {1986})}\BibitemShut {NoStop}%
\bibitem [{\citenamefont {Bierhorst}(2015)}]{B15}%
  \BibitemOpen
  \bibfield  {author} {\bibinfo {author} {\bibfnamefont {P.}~\bibnamefont
  {Bierhorst}},\ }\href {http://stacks.iop.org/1751-8121/48/i=19/a=195302}
  {\bibfield  {journal} {\bibinfo  {journal} {Journal of Physics A:
  Mathematical and Theoretical}\ }\textbf {\bibinfo {volume} {48}},\ \bibinfo
  {pages} {195302} (\bibinfo {year} {2015})}\BibitemShut {NoStop}%
\bibitem [{Note2()}]{Note2}%
  \BibitemOpen
  \bibinfo {note} {In the case they would rather use several pairs of inputs
  for randomness generation, the analysis below could be extended by using the
  tools presented in~\cite {bancal13more}.}\BibitemShut {Stop}%
\bibitem [{\citenamefont {Shaltiel}()}]{shaltiel2002recent}%
  \BibitemOpen
  \bibfield  {author} {\bibinfo {author} {\bibfnamefont {R.}~\bibnamefont
  {Shaltiel}},\ }\href@noop {} {\bibfield  {journal} {\bibinfo  {journal}
  {Bulletin of the EATCS}\ }\textbf {\bibinfo {volume} {77}},\ \bibinfo {pages}
  {10}}\BibitemShut {NoStop}%
\bibitem [{\citenamefont {De}\ \emph {et~al.}(2012)\citenamefont {De},
  \citenamefont {Portmann}, \citenamefont {Vidick},\ and\ \citenamefont
  {Renner}}]{DPVR12}%
  \BibitemOpen
  \bibfield  {author} {\bibinfo {author} {\bibfnamefont {A.}~\bibnamefont
  {De}}, \bibinfo {author} {\bibfnamefont {C.}~\bibnamefont {Portmann}},
  \bibinfo {author} {\bibfnamefont {T.}~\bibnamefont {Vidick}}, \ and\ \bibinfo
  {author} {\bibfnamefont {R.}~\bibnamefont {Renner}},\ }\href {\doibase
  10.1137/100813683} {\bibfield  {journal} {\bibinfo  {journal} {SIAM Journal
  on Computing}\ }\textbf {\bibinfo {volume} {41}},\ \bibinfo {pages} {915}
  (\bibinfo {year} {2012})},\ \Eprint
  {http://arxiv.org/abs/http://dx.doi.org/10.1137/100813683}
  {http://dx.doi.org/10.1137/100813683} \BibitemShut {NoStop}%
\bibitem [{Note3()}]{Note3}%
  \BibitemOpen
  \bibinfo {note} {Note that usually, the average min-entropy of a variable $A$
  (e.g. the output string) given some information $M$ (e.g. the length of the
  post-selected string) is defined as $-\protect \qopname \relax o{log}_2
  \DOTSB \sum@ \slimits@ _m P(m) G(A|M=m)$ where $G(A|M=m)$ is the guessing
  probability of $A$ given $M=m$. Here our definition of ``average''
  min-entropy is $\DOTSB \sum@ \slimits@ _m P(m)(-\protect \qopname \relax
  o{log}_2 G(A|M=m))$, that is, we inverted the sum over $m$ and the logarithm.
  The reason is that in our scenario, the user, and not only the adversary,
  actually knows the value of $m$ and thus a bound on $G(A|M=m)$ which allows
  him to apply a different extractor depending on $m$.}\BibitemShut {Stop}%
\bibitem [{\citenamefont {Konig}\ \emph {et~al.}(2009)\citenamefont {Konig},
  \citenamefont {Renner},\ and\ \citenamefont {Schaffner}}]{KRS09}%
  \BibitemOpen
  \bibfield  {author} {\bibinfo {author} {\bibfnamefont {R.}~\bibnamefont
  {Konig}}, \bibinfo {author} {\bibfnamefont {R.}~\bibnamefont {Renner}}, \
  and\ \bibinfo {author} {\bibfnamefont {C.}~\bibnamefont {Schaffner}},\ }\href
  {\doibase 10.1109/TIT.2009.2025545} {\bibfield  {journal} {\bibinfo
  {journal} {Information Theory, IEEE Transactions on}\ }\textbf {\bibinfo
  {volume} {55}},\ \bibinfo {pages} {4337} (\bibinfo {year}
  {2009})}\BibitemShut {NoStop}%
\bibitem [{\citenamefont {Navascu{\'e}s}\ \emph {et~al.}(2007)\citenamefont
  {Navascu{\'e}s}, \citenamefont {Pironio},\ and\ \citenamefont
  {Ac{\'\i}n}}]{qtest0}%
  \BibitemOpen
  \bibfield  {author} {\bibinfo {author} {\bibfnamefont {M.}~\bibnamefont
  {Navascu{\'e}s}}, \bibinfo {author} {\bibfnamefont {S.}~\bibnamefont
  {Pironio}}, \ and\ \bibinfo {author} {\bibfnamefont {A.}~\bibnamefont
  {Ac{\'\i}n}},\ }\href@noop {} {\bibfield  {journal} {\bibinfo  {journal}
  {Phys. Rev. Lett.}\ }\textbf {\bibinfo {volume} {98}},\ \bibinfo {pages}
  {010401} (\bibinfo {year} {2007})}\BibitemShut {NoStop}%
\bibitem [{\citenamefont {Navascu{\'e}s}\ \emph {et~al.}(2008)\citenamefont
  {Navascu{\'e}s}, \citenamefont {Pironio},\ and\ \citenamefont
  {Ac{\'\i}n}}]{qtest}%
  \BibitemOpen
  \bibfield  {author} {\bibinfo {author} {\bibfnamefont {M.}~\bibnamefont
  {Navascu{\'e}s}}, \bibinfo {author} {\bibfnamefont {S.}~\bibnamefont
  {Pironio}}, \ and\ \bibinfo {author} {\bibfnamefont {A.}~\bibnamefont
  {Ac{\'\i}n}},\ }\href@noop {} {\bibfield  {journal} {\bibinfo  {journal} {New
  J. Phys.}\ }\textbf {\bibinfo {volume} {10}},\ \bibinfo {pages} {073013}
  (\bibinfo {year} {2008})}\BibitemShut {NoStop}%
\bibitem [{\citenamefont {Pironio}\ \emph
  {et~al.}(2010{\natexlab{b}})\citenamefont {Pironio}, \citenamefont
  {Navascu{\'e}s},\ and\ \citenamefont {Ac{\'\i}n}}]{qtest2}%
  \BibitemOpen
  \bibfield  {author} {\bibinfo {author} {\bibfnamefont {S.}~\bibnamefont
  {Pironio}}, \bibinfo {author} {\bibfnamefont {M.}~\bibnamefont
  {Navascu{\'e}s}}, \ and\ \bibinfo {author} {\bibfnamefont {A.}~\bibnamefont
  {Ac{\'\i}n}},\ }\href@noop {} {\bibfield  {journal} {\bibinfo  {journal}
  {SIAM J. Optim.}\ }\textbf {\bibinfo {volume} {20}},\ \bibinfo {pages} {2157}
  (\bibinfo {year} {2010}{\natexlab{b}})}\BibitemShut {NoStop}%
\bibitem [{\citenamefont {Moroder}\ \emph {et~al.}(2013)\citenamefont
  {Moroder}, \citenamefont {Bancal}, \citenamefont {Liang}, \citenamefont
  {Hofmann},\ and\ \citenamefont {G\"uhne}}]{moroder2013}%
  \BibitemOpen
  \bibfield  {author} {\bibinfo {author} {\bibfnamefont {T.}~\bibnamefont
  {Moroder}}, \bibinfo {author} {\bibfnamefont {J.-D.}\ \bibnamefont {Bancal}},
  \bibinfo {author} {\bibfnamefont {Y.-C.}\ \bibnamefont {Liang}}, \bibinfo
  {author} {\bibfnamefont {M.}~\bibnamefont {Hofmann}}, \ and\ \bibinfo
  {author} {\bibfnamefont {O.}~\bibnamefont {G\"uhne}},\ }\href {\doibase
  10.1103/PhysRevLett.111.030501} {\bibfield  {journal} {\bibinfo  {journal}
  {Phys. Rev. Lett.}\ }\textbf {\bibinfo {volume} {111}},\ \bibinfo {pages}
  {030501} (\bibinfo {year} {2013})}\BibitemShut {NoStop}%
\bibitem [{\citenamefont {Caprara~Vivoli}\ \emph {et~al.}(2015)\citenamefont
  {Caprara~Vivoli}, \citenamefont {Sekatski}, \citenamefont {Bancal},
  \citenamefont {Lim}, \citenamefont {Christensen}, \citenamefont {Martin},
  \citenamefont {Thew}, \citenamefont {Zbinden}, \citenamefont {Gisin},\ and\
  \citenamefont {Sangouard}}]{VSBLCMTZGS15}%
  \BibitemOpen
  \bibfield  {author} {\bibinfo {author} {\bibfnamefont {V.}~\bibnamefont
  {Caprara~Vivoli}}, \bibinfo {author} {\bibfnamefont {P.}~\bibnamefont
  {Sekatski}}, \bibinfo {author} {\bibfnamefont {J.-D.}\ \bibnamefont
  {Bancal}}, \bibinfo {author} {\bibfnamefont {C.~C.~W.}\ \bibnamefont {Lim}},
  \bibinfo {author} {\bibfnamefont {B.~G.}\ \bibnamefont {Christensen}},
  \bibinfo {author} {\bibfnamefont {A.}~\bibnamefont {Martin}}, \bibinfo
  {author} {\bibfnamefont {R.~T.}\ \bibnamefont {Thew}}, \bibinfo {author}
  {\bibfnamefont {H.}~\bibnamefont {Zbinden}}, \bibinfo {author} {\bibfnamefont
  {N.}~\bibnamefont {Gisin}}, \ and\ \bibinfo {author} {\bibfnamefont
  {N.}~\bibnamefont {Sangouard}},\ }\href {\doibase 10.1103/PhysRevA.91.012107}
  {\bibfield  {journal} {\bibinfo  {journal} {Phys. Rev. A}\ }\textbf {\bibinfo
  {volume} {91}},\ \bibinfo {pages} {012107} (\bibinfo {year}
  {2015})}\BibitemShut {NoStop}%
\bibitem [{\citenamefont {Hensen}\ \emph {et~al.}(2015)\citenamefont {Hensen},
  \citenamefont {Bernien}, \citenamefont {Dreau}, \citenamefont {Reiserer},
  \citenamefont {Kalb}, \citenamefont {Blok}, \citenamefont {Ruitenberg},
  \citenamefont {Vermeulen}, \citenamefont {Schouten}, \citenamefont {Abellan},
  \citenamefont {Amaya}, \citenamefont {Pruneri}, \citenamefont {Mitchell},
  \citenamefont {Markham}, \citenamefont {Twitchen}, \citenamefont {Elkouss},
  \citenamefont {Wehner}, \citenamefont {Taminiau},\ and\ \citenamefont
  {Hanson}}]{HBDRKBRVSAAPMMTEWTH15}%
  \BibitemOpen
  \bibfield  {author} {\bibinfo {author} {\bibfnamefont {B.}~\bibnamefont
  {Hensen}}, \bibinfo {author} {\bibfnamefont {H.}~\bibnamefont {Bernien}},
  \bibinfo {author} {\bibfnamefont {A.~E.}\ \bibnamefont {Dreau}}, \bibinfo
  {author} {\bibfnamefont {A.}~\bibnamefont {Reiserer}}, \bibinfo {author}
  {\bibfnamefont {N.}~\bibnamefont {Kalb}}, \bibinfo {author} {\bibfnamefont
  {M.~S.}\ \bibnamefont {Blok}}, \bibinfo {author} {\bibfnamefont
  {J.}~\bibnamefont {Ruitenberg}}, \bibinfo {author} {\bibfnamefont {R.~F.~L.}\
  \bibnamefont {Vermeulen}}, \bibinfo {author} {\bibfnamefont {R.~N.}\
  \bibnamefont {Schouten}}, \bibinfo {author} {\bibfnamefont {C.}~\bibnamefont
  {Abellan}}, \bibinfo {author} {\bibfnamefont {W.}~\bibnamefont {Amaya}},
  \bibinfo {author} {\bibfnamefont {V.}~\bibnamefont {Pruneri}}, \bibinfo
  {author} {\bibfnamefont {M.~W.}\ \bibnamefont {Mitchell}}, \bibinfo {author}
  {\bibfnamefont {M.}~\bibnamefont {Markham}}, \bibinfo {author} {\bibfnamefont
  {D.~J.}\ \bibnamefont {Twitchen}}, \bibinfo {author} {\bibfnamefont
  {D.}~\bibnamefont {Elkouss}}, \bibinfo {author} {\bibfnamefont
  {S.}~\bibnamefont {Wehner}}, \bibinfo {author} {\bibfnamefont {T.~H.}\
  \bibnamefont {Taminiau}}, \ and\ \bibinfo {author} {\bibfnamefont
  {R.}~\bibnamefont {Hanson}},\ }\href@noop {} {\bibfield  {journal} {\bibinfo
  {journal} {Nature}\ }\textbf {\bibinfo {volume} {526}},\ \bibinfo {pages}
  {682} (\bibinfo {year} {2015})}\BibitemShut {NoStop}%
\bibitem [{\citenamefont {M\'attar}\ \emph {et~al.}(2015)\citenamefont
  {M\'attar}, \citenamefont {Skrzypczyk}, \citenamefont {Brask}, \citenamefont
  {Cavalcanti},\ and\ \citenamefont {Ac\'in}}]{MSBCA15}%
  \BibitemOpen
  \bibfield  {author} {\bibinfo {author} {\bibfnamefont {A.}~\bibnamefont
  {M\'attar}}, \bibinfo {author} {\bibfnamefont {P.}~\bibnamefont
  {Skrzypczyk}}, \bibinfo {author} {\bibfnamefont {J.~B.}\ \bibnamefont
  {Brask}}, \bibinfo {author} {\bibfnamefont {D.}~\bibnamefont {Cavalcanti}}, \
  and\ \bibinfo {author} {\bibfnamefont {A.}~\bibnamefont {Ac\'in}},\ }\href
  {http://stacks.iop.org/1367-2630/17/i=2/a=022003} {\bibfield  {journal}
  {\bibinfo  {journal} {New Journal of Physics}\ }\textbf {\bibinfo {volume}
  {17}},\ \bibinfo {pages} {022003} (\bibinfo {year} {2015})}\BibitemShut
  {NoStop}%
\bibitem [{\citenamefont {Shalm}\ \emph {et~al.}(2015)\citenamefont {Shalm},
  \citenamefont {Meyer-Scott}, \citenamefont {Christensen}, \citenamefont
  {Bierhorst}, \citenamefont {Wayne}, \citenamefont {Stevens}, \citenamefont
  {Gerrits}, \citenamefont {Glancy}, \citenamefont {Hamel}, \citenamefont
  {Allman}, \citenamefont {Coakley}, \citenamefont {Dyer}, \citenamefont
  {Hodge}, \citenamefont {Lita}, \citenamefont {Verma}, \citenamefont
  {Lambrocco}, \citenamefont {Tortorici}, \citenamefont {Migdall},
  \citenamefont {Zhang}, \citenamefont {Kumor}, \citenamefont {Farr},
  \citenamefont {Marsili}, \citenamefont {Shaw}, \citenamefont {Stern},
  \citenamefont {Abell\'an}, \citenamefont {Amaya}, \citenamefont {Pruneri},
  \citenamefont {Jennewein}, \citenamefont {Mitchell}, \citenamefont {Kwiat},
  \citenamefont {Bienfang}, \citenamefont {Mirin}, \citenamefont {Knill},\ and\
  \citenamefont {Nam}}]{Shalm15}%
  \BibitemOpen
  \bibfield  {author} {\bibinfo {author} {\bibfnamefont {L.~K.}\ \bibnamefont
  {Shalm}}, \bibinfo {author} {\bibfnamefont {E.}~\bibnamefont {Meyer-Scott}},
  \bibinfo {author} {\bibfnamefont {B.~G.}\ \bibnamefont {Christensen}},
  \bibinfo {author} {\bibfnamefont {P.}~\bibnamefont {Bierhorst}}, \bibinfo
  {author} {\bibfnamefont {M.~A.}\ \bibnamefont {Wayne}}, \bibinfo {author}
  {\bibfnamefont {M.~J.}\ \bibnamefont {Stevens}}, \bibinfo {author}
  {\bibfnamefont {T.}~\bibnamefont {Gerrits}}, \bibinfo {author} {\bibfnamefont
  {S.}~\bibnamefont {Glancy}}, \bibinfo {author} {\bibfnamefont {D.~R.}\
  \bibnamefont {Hamel}}, \bibinfo {author} {\bibfnamefont {M.~S.}\ \bibnamefont
  {Allman}}, \bibinfo {author} {\bibfnamefont {K.~J.}\ \bibnamefont {Coakley}},
  \bibinfo {author} {\bibfnamefont {S.~D.}\ \bibnamefont {Dyer}}, \bibinfo
  {author} {\bibfnamefont {C.}~\bibnamefont {Hodge}}, \bibinfo {author}
  {\bibfnamefont {A.~E.}\ \bibnamefont {Lita}}, \bibinfo {author}
  {\bibfnamefont {V.~B.}\ \bibnamefont {Verma}}, \bibinfo {author}
  {\bibfnamefont {C.}~\bibnamefont {Lambrocco}}, \bibinfo {author}
  {\bibfnamefont {E.}~\bibnamefont {Tortorici}}, \bibinfo {author}
  {\bibfnamefont {A.~L.}\ \bibnamefont {Migdall}}, \bibinfo {author}
  {\bibfnamefont {Y.}~\bibnamefont {Zhang}}, \bibinfo {author} {\bibfnamefont
  {D.~R.}\ \bibnamefont {Kumor}}, \bibinfo {author} {\bibfnamefont {W.~H.}\
  \bibnamefont {Farr}}, \bibinfo {author} {\bibfnamefont {F.}~\bibnamefont
  {Marsili}}, \bibinfo {author} {\bibfnamefont {M.~D.}\ \bibnamefont {Shaw}},
  \bibinfo {author} {\bibfnamefont {J.~A.}\ \bibnamefont {Stern}}, \bibinfo
  {author} {\bibfnamefont {C.}~\bibnamefont {Abell\'an}}, \bibinfo {author}
  {\bibfnamefont {W.}~\bibnamefont {Amaya}}, \bibinfo {author} {\bibfnamefont
  {V.}~\bibnamefont {Pruneri}}, \bibinfo {author} {\bibfnamefont
  {T.}~\bibnamefont {Jennewein}}, \bibinfo {author} {\bibfnamefont {M.~W.}\
  \bibnamefont {Mitchell}}, \bibinfo {author} {\bibfnamefont {P.~G.}\
  \bibnamefont {Kwiat}}, \bibinfo {author} {\bibfnamefont {J.~C.}\ \bibnamefont
  {Bienfang}}, \bibinfo {author} {\bibfnamefont {R.~P.}\ \bibnamefont {Mirin}},
  \bibinfo {author} {\bibfnamefont {E.}~\bibnamefont {Knill}}, \ and\ \bibinfo
  {author} {\bibfnamefont {S.~W.}\ \bibnamefont {Nam}},\ }\href {\doibase
  10.1103/PhysRevLett.115.250402} {\bibfield  {journal} {\bibinfo  {journal}
  {Phys. Rev. Lett.}\ }\textbf {\bibinfo {volume} {115}},\ \bibinfo {pages}
  {250402} (\bibinfo {year} {2015})}\BibitemShut {NoStop}%
\bibitem [{\citenamefont {Giustina}\ \emph {et~al.}(2015)\citenamefont
  {Giustina}, \citenamefont {Versteegh}, \citenamefont {Wengerowsky},
  \citenamefont {Handsteiner}, \citenamefont {Hochrainer}, \citenamefont
  {Phelan}, \citenamefont {Steinlechner}, \citenamefont {Kofler}, \citenamefont
  {Larsson}, \citenamefont {Abell\'an}, \citenamefont {Amaya}, \citenamefont
  {Pruneri}, \citenamefont {Mitchell}, \citenamefont {Beyer}, \citenamefont
  {Gerrits}, \citenamefont {Lita}, \citenamefont {Shalm}, \citenamefont {Nam},
  \citenamefont {Scheidl}, \citenamefont {Ursin}, \citenamefont {Wittmann},\
  and\ \citenamefont {Zeilinger}}]{Giustina15}%
  \BibitemOpen
  \bibfield  {author} {\bibinfo {author} {\bibfnamefont {M.}~\bibnamefont
  {Giustina}}, \bibinfo {author} {\bibfnamefont {M.~A.~M.}\ \bibnamefont
  {Versteegh}}, \bibinfo {author} {\bibfnamefont {S.}~\bibnamefont
  {Wengerowsky}}, \bibinfo {author} {\bibfnamefont {J.}~\bibnamefont
  {Handsteiner}}, \bibinfo {author} {\bibfnamefont {A.}~\bibnamefont
  {Hochrainer}}, \bibinfo {author} {\bibfnamefont {K.}~\bibnamefont {Phelan}},
  \bibinfo {author} {\bibfnamefont {F.}~\bibnamefont {Steinlechner}}, \bibinfo
  {author} {\bibfnamefont {J.}~\bibnamefont {Kofler}}, \bibinfo {author}
  {\bibfnamefont {J.-A.}\ \bibnamefont {Larsson}}, \bibinfo {author}
  {\bibfnamefont {C.}~\bibnamefont {Abell\'an}}, \bibinfo {author}
  {\bibfnamefont {W.}~\bibnamefont {Amaya}}, \bibinfo {author} {\bibfnamefont
  {V.}~\bibnamefont {Pruneri}}, \bibinfo {author} {\bibfnamefont {M.~W.}\
  \bibnamefont {Mitchell}}, \bibinfo {author} {\bibfnamefont {J.}~\bibnamefont
  {Beyer}}, \bibinfo {author} {\bibfnamefont {T.}~\bibnamefont {Gerrits}},
  \bibinfo {author} {\bibfnamefont {A.~E.}\ \bibnamefont {Lita}}, \bibinfo
  {author} {\bibfnamefont {L.~K.}\ \bibnamefont {Shalm}}, \bibinfo {author}
  {\bibfnamefont {S.~W.}\ \bibnamefont {Nam}}, \bibinfo {author} {\bibfnamefont
  {T.}~\bibnamefont {Scheidl}}, \bibinfo {author} {\bibfnamefont
  {R.}~\bibnamefont {Ursin}}, \bibinfo {author} {\bibfnamefont
  {B.}~\bibnamefont {Wittmann}}, \ and\ \bibinfo {author} {\bibfnamefont
  {A.}~\bibnamefont {Zeilinger}},\ }\href {\doibase
  10.1103/PhysRevLett.115.250401} {\bibfield  {journal} {\bibinfo  {journal}
  {Phys. Rev. Lett.}\ }\textbf {\bibinfo {volume} {115}},\ \bibinfo {pages}
  {250401} (\bibinfo {year} {2015})}\BibitemShut {NoStop}%
\bibitem [{\citenamefont {Massar}\ \emph {et~al.}(2002)\citenamefont {Massar},
  \citenamefont {Pironio}, \citenamefont {Roland},\ and\ \citenamefont
  {Gisin}}]{MPRG02}%
  \BibitemOpen
  \bibfield  {author} {\bibinfo {author} {\bibfnamefont {S.}~\bibnamefont
  {Massar}}, \bibinfo {author} {\bibfnamefont {S.}~\bibnamefont {Pironio}},
  \bibinfo {author} {\bibfnamefont {J.}~\bibnamefont {Roland}}, \ and\ \bibinfo
  {author} {\bibfnamefont {B.}~\bibnamefont {Gisin}},\ }\href {\doibase
  10.1103/PhysRevA.66.052112} {\bibfield  {journal} {\bibinfo  {journal} {Phys.
  Rev. A}\ }\textbf {\bibinfo {volume} {66}},\ \bibinfo {pages} {052112}
  (\bibinfo {year} {2002})}\BibitemShut {NoStop}%
\bibitem [{\citenamefont {Pironio}(2005)}]{P05lifting}%
  \BibitemOpen
  \bibfield  {author} {\bibinfo {author} {\bibfnamefont {S.}~\bibnamefont
  {Pironio}},\ }\href {\doibase http://dx.doi.org/10.1063/1.1928727} {\bibfield
   {journal} {\bibinfo  {journal} {Journal of Mathematical Physics}\ }\textbf
  {\bibinfo {volume} {46}},\ \bibinfo {eid} {062112} (\bibinfo {year}
  {2005})}\BibitemShut {NoStop}%
\bibitem [{\citenamefont {Barrett}\ \emph {et~al.}(2002)\citenamefont
  {Barrett}, \citenamefont {Collins}, \citenamefont {Hardy}, \citenamefont
  {Kent},\ and\ \citenamefont {Popescu}}]{BCHKP02}%
  \BibitemOpen
  \bibfield  {author} {\bibinfo {author} {\bibfnamefont {J.}~\bibnamefont
  {Barrett}}, \bibinfo {author} {\bibfnamefont {D.}~\bibnamefont {Collins}},
  \bibinfo {author} {\bibfnamefont {L.}~\bibnamefont {Hardy}}, \bibinfo
  {author} {\bibfnamefont {A.}~\bibnamefont {Kent}}, \ and\ \bibinfo {author}
  {\bibfnamefont {S.}~\bibnamefont {Popescu}},\ }\href {\doibase
  10.1103/PhysRevA.66.042111} {\bibfield  {journal} {\bibinfo  {journal} {Phys.
  Rev. A}\ }\textbf {\bibinfo {volume} {66}},\ \bibinfo {pages} {042111}
  (\bibinfo {year} {2002})}\BibitemShut {NoStop}%
\bibitem [{\citenamefont {Gill}(2014)}]{gill14}%
  \BibitemOpen
  \bibfield  {author} {\bibinfo {author} {\bibfnamefont {R.~D.}\ \bibnamefont
  {Gill}},\ }\href {\doibase 10.1214/14-STS490} {\bibfield  {journal} {\bibinfo
   {journal} {Statist. Sci.}\ }\textbf {\bibinfo {volume} {29}},\ \bibinfo
  {pages} {512} (\bibinfo {year} {2014})}\BibitemShut {NoStop}%
\bibitem [{\citenamefont {{Renner}}(2005)}]{rennerThesis}%
  \BibitemOpen
  \bibfield  {author} {\bibinfo {author} {\bibfnamefont {R.}~\bibnamefont
  {{Renner}}},\ }\emph {\bibinfo {title} {{Security of Quantum Key
  Distribution}}},\ \href@noop {} {Ph.D. thesis},\ \bibinfo  {school} {PhD
  Thesis, 2005} (\bibinfo {year} {2005})\BibitemShut {NoStop}%
\bibitem [{\citenamefont {Barrett}\ \emph {et~al.}(2013)\citenamefont
  {Barrett}, \citenamefont {Colbeck},\ and\ \citenamefont {Kent}}]{BCK13}%
  \BibitemOpen
  \bibfield  {author} {\bibinfo {author} {\bibfnamefont {J.}~\bibnamefont
  {Barrett}}, \bibinfo {author} {\bibfnamefont {R.}~\bibnamefont {Colbeck}}, \
  and\ \bibinfo {author} {\bibfnamefont {A.}~\bibnamefont {Kent}},\ }\href
  {\doibase 10.1103/PhysRevLett.110.010503} {\bibfield  {journal} {\bibinfo
  {journal} {Phys. Rev. Lett.}\ }\textbf {\bibinfo {volume} {110}},\ \bibinfo
  {pages} {010503} (\bibinfo {year} {2013})}\BibitemShut {NoStop}%
\bibitem [{\citenamefont {Law}\ \emph {et~al.}(2014)\citenamefont {Law},
  \citenamefont {Thinh}, \citenamefont {Bancal},\ and\ \citenamefont
  {Scarani}}]{LTBS14}%
  \BibitemOpen
  \bibfield  {author} {\bibinfo {author} {\bibfnamefont {Y.~Z.}\ \bibnamefont
  {Law}}, \bibinfo {author} {\bibfnamefont {L.~P.}\ \bibnamefont {Thinh}},
  \bibinfo {author} {\bibfnamefont {J.-D.}\ \bibnamefont {Bancal}}, \ and\
  \bibinfo {author} {\bibfnamefont {V.}~\bibnamefont {Scarani}},\ }\href
  {http://stacks.iop.org/1751-8121/47/i=42/a=424028} {\bibfield  {journal}
  {\bibinfo  {journal} {Journal of Physics A: Mathematical and Theoretical}\
  }\textbf {\bibinfo {volume} {47}},\ \bibinfo {pages} {424028} (\bibinfo
  {year} {2014})}\BibitemShut {NoStop}%
\bibitem [{\citenamefont {Collins}\ and\ \citenamefont {Gisin}(2004)}]{CG04}%
  \BibitemOpen
  \bibfield  {author} {\bibinfo {author} {\bibfnamefont {D.}~\bibnamefont
  {Collins}}\ and\ \bibinfo {author} {\bibfnamefont {N.}~\bibnamefont
  {Gisin}},\ }\href {http://stacks.iop.org/0305-4470/37/i=5/a=021} {\bibfield
  {journal} {\bibinfo  {journal} {Journal of Physics A: Mathematical and
  General}\ }\textbf {\bibinfo {volume} {37}},\ \bibinfo {pages} {1775}
  (\bibinfo {year} {2004})}\BibitemShut {NoStop}%
\end{thebibliography}%

\end{document}